\documentclass[
reprint,
groupedaddress,
superscriptaddress,
nobibnotes,
secnumarabic,
amsmath,amssymb,
prb,
]{revtex4-1}

\usepackage{amsmath}
\usepackage{graphicx}
\usepackage{bm}
\usepackage[utf8]{inputenc} 
\usepackage{hyperref}
\usepackage{xcolor}
\newcommand{\red}{\textcolor{black}}

\newcommand{\mos}{MoS$_2$}

\newcommand{\gm}{|\Gamma M|}
\makeatletter
\newcommand*{\rom}[1]{\expandafter\@slowromancap\romannumeral #1@}
\makeatother
\begin{document}


\title{Flat-band plasmons in twisted bilayer transition metal dichalcogenides}

\author{Xueheng Kuang}
\affiliation{Key Laboratory of Artificial Micro- and Nano-structures of Ministry of Education and School of Physics and Technology, Wuhan University, Wuhan 430072, China}
\affiliation{Imdea Nanoscience, Faraday 9, 28015 Madrid, Spain}
\author{Zhen Zhan}
\email{Corresponding author: zhen.zhan@whu.edu.cn}
\affiliation{Key Laboratory of Artificial Micro- and Nano-structures of Ministry of Education and School of Physics and Technology, Wuhan University, Wuhan 430072, China}

\author{Shengjun Yuan}
\email{Corresponding author: s.yuan@whu.edu.cn}
\affiliation{Key Laboratory of Artificial Micro- and Nano-structures of Ministry of Education and School of Physics and Technology, Wuhan University, Wuhan 430072, China}
\affiliation{Wuhan Institute of Quantum Technology, Wuhan 430206, China}

\date{\today}

\begin{abstract}
Twisted bilayer transition metal dichalcogenides are ideal platforms to study flat-band phenomena. In this paper, we investigate flat-band plasmons in \red{the} hole-doped twisted bilayer MoS$_2$ \red{(tb-MoS$_2$)} by em­ploying a full tight-binding model and the random phase approximation. \red{When considering lattice relaxations in tb-MoS$_2$}, the flat band is not separated from remote valence bands, which makes the contribution of interband transitions in transforming \red{the} plasmon dispersion and energy significantly different. In particular, low-damped and quasi-flat plasmons emerge if we only consider intraband transitions in the doped flat band, whereas a $\sqrt q$ plasmon dispersion emerges if we also take into account interband transitions between the flat band and remote bands. Furthermore, the plasmon energies are tunable with twist angle\red{s} and doping level\red{s}. However, in a rigid sample \red{that suffers no lattice relaxations}, lower-energy quasi-flat plasmons and higher-energy interband plasmons can coexist. \red{For rigid tb-MoS$_2$ with a high doping level,} strongly enhanced interband transitions quench the quasi-flat plasmons. Based on the \red{lattice} relaxation and doping effects, we conclude that two conditions, \red{the} isolated flat band and \red{a} properly hole-doping level, are essential for observing the low-damped and quasi-flat plasmon mode in twisted bilayer transi­tion metal dichalcogenides. We hope that our study on flat-band plasmons can be instructive for studying the possibility of plasmon-mediated superconductivity in twisted bilayer transition metal dichalcogenides in the future.
 
\end{abstract}


\maketitle

\section{introduction}

Twisted bilayer graphene (TBG) with flat bands has opened an avenue to explore abundant phenomena, for instance, localized and correlated states\cite{pnas2011moire,cao2018correlated,nature2019fulltb}, unconventional superconductivity\cite{cao2018unconventional,science2019tuning}, and electronic collective excitations\cite{plasexp2019cao,pnasplas2019intrinsically,Kuang2021}. Collective excited modes arising from quasi-localized states of flat bands, named as flat-band plasmons, feature intrinsically undamped behaviors and constant energy dispersion\cite{pnasplas2019intrinsically,Kuang2021}, giving insight into the unconventional superconductivity\cite{prs2020superconductivity,lewandowski2020pairing,tbg2021plassuper} and linear resistivity experimentally observed in TBG\cite{marginal}. Recently, ultraflat bands are detected in twisted bilayer transition metal dichalcogenides (tb-TMDs) with a wide range of angles\cite{tb-mos2018banddft, Zhan2020,flatband2019visualization,flatband2021tb-tmd,Zhang2020,flat2021wsesoc}, making tb-TMDs ideal platforms to extensively investigate many-body states\cite{tb-wse2020correlat,tb-wse2021manybody,tb-mos2021kpmodel, tb-wse2021manybody,collec2020tbmos2,exptb-wse2019corrlatesuper,exptb-wse2020correlate,exptb-wse2020flatband,exptb-wse2021flatband2} and optical excitons\cite{tb-tmd2021exciton,exptb-mos2021optical,tb-tmd2019excition}. For example, zero-resistance pockets are observed on doping away from half filling of the flat band in twisted bilayer WSe$_2$, which indicates a possible transition to a superconducting state\cite{exptb-wse2019corrlatesuper}. Theoretical studies establish that heterobilayer transition metal dichalcogenides are unique platforms to realize chiral superconductivity\cite{fuliang2021supertmd,chiral2021super}. Potential superconducting parings arising from magnon and spin-valley fluctuations are proposed in tb-TMDs\cite{magnon2022super, fuliang2021supertmd}. 

Previous studies show that plasmon properties play a role in the pairing interaction responsible for superconductivity in TBG\cite{prs2020superconductivity, lewandowski2020pairing,tbg2021plassuper}. The plasmon-mediated superconductivity is determined by a ratio of \red{the} plasmon energy to \red{the} flat-band bandwidth. That is, with the flat-band plasmon energy scale comparable to the flat-band bandwidth\cite{pnasplas2019intrinsically, Kuang2021}, a superconducting state can be realized in TBG\cite{prs2020superconductivity, lewandowski2020pairing}. Then, we wonder whether plasmons in flat-band tb-TMDs possess similar properties as that in TBG and could contribute to parings in tb-TMDs. 
	Up to now, plasmonic properties of flat-band tb-TMDs are still not clear, which hinders us from further studying the plasmon-mediated superconductivity. The presence of flat bands in tb-{\mos} may result in different plasmon properties from those surveyed in monolayer\cite{1Lmos2013plasstauber,1Lmosacoustic2014,1lmos2016vallyplas,1Lmos2017plas_acoustic,1lmos2017nonlocalplas,review2019plas2D}, two-layer\cite{2Lmos2017plasacoustic}, few-layer\cite{few-Lmos2017plasexciton,1L-mos2020plasdft}, and one-sheet {\mos} systems\cite{sheetmos2014plas,1lmos2021plas-optic}, since the unique flat-band plasmons detected in TBG are distinct from those discovered in monolayer and bilayer graphene\cite{monogra2007dielectric,tonylow2014novel}. In practice, such \red{a} unique flat-band plasmon with undamped and quasi-flat characteristics can also lead to special applications such as photon-based quantum information processing toolbox and perfect lens\cite{pnasplas2019intrinsically,nano2016plas}. 
	All in all, the property of plasmons in flat-band tb-TMDs deserves further investigation.

In this paper, we mainly focus on flat-band plasmons in twisted bilayer {\mos} (tb-{\mos}). Previous studies show that the tb-{\mos} systems are semiconductors with ultra-flat bands in the valence band maximum (VBM). The flat bands are discovered in tb-{\mos} with a wide range of twist angles and have narrower bandwidth at a smaller angle\cite{tb-mos2018banddft,flatband2021tb-tmd,isoband2020tbmos2}. After introduce hole doping in the VBM, we employ a full tight-binding (TB) model to investigate low-energy plasmons in tb-{\mos}. It is known that the bandwidth of \red{a} flat band obviously modulates plasmon properties in TBG\cite{nano2016plas,Kuang2021}. By changing the twist angle of tb-{\mos}, we can also study how flatness of the flat band modifies the collective excitations. Moreover, the lattice relaxation significantly changes the electronic properties of tb-TMDs \red{with} small twist angles\cite{Zhan2020}. With lattice relaxation considered in tb-{\mos}, the band gap between the flat \red{VBM} and other valence bands disappears at large twist angles\cite{tb-mos2018banddft,flatband2021tb-tmd}. Will the absence of the band gap affect the flat-band plasmon?  In principle, the polarization function can be calculated via the Lindhard function\cite{linear2005quantum}. With this method, we can investigate the effect of band cutoff\red{s} on the flat-band plasmon. For example, we can perform a one-band calculation where only flat-band intraband transitions contribute to the flat-band plasmon. Meanwhile, a full-band calculation can also be realized via a combination of the Kubo formula and  tight-binding propagation method (TBPM)\cite{yuan2010tipsi,yuan2011kubo}. In the full-band calculation, both intraband and interband polarizations are taken into account. Therefore, interband transition effects on plasmons \red{in tb-{\mos}} are investigated by comparing the modes obtained from the one-band and full-band calculations. Our work could be an example to study flat-band plasmons in other twisted 2D semiconductors. 

This paper is organized as follows. In  Sec. II, the tight-binding model and computational methods are introduced. In Sec. III and Sec. IV, flat-band plasmons are explicitly studied in both relaxed (consider the atomic relaxation) and rigid (without \red{the} atomic relaxation) tb-{\mos}, respectively. In Sec. V, we pay attention to the effects of band cutoff\red{s} and chemical potentials on plasmons. Finally, we give a summary and discussion of our work.

\section{Numerical methods}
\label{sec2}

\begin{figure}[t]	
	\includegraphics[width=0.45\textwidth]{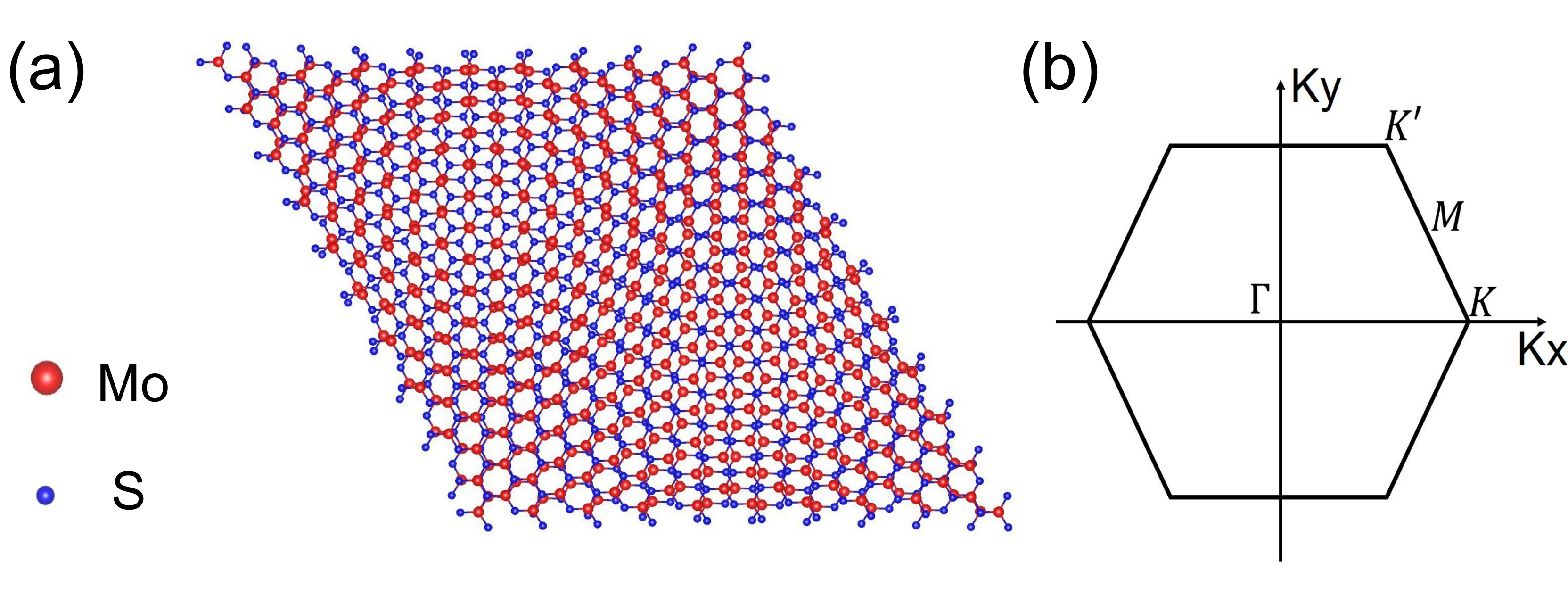}
	\caption{(a) Top view of the atomic structure of $3.5^\circ$ tb-{\mos}. (b) The first BZ with high symmetry points for the hexagonal lattice of tb-{\mos}.}
	\label{fig:structre}
\end{figure}

\subsection{Tight-binding model}
We construct atomic structures of tb-{\mos} with a commensurate approach used in building TBG structures\cite{structure2016universal,strustrain2018prl}. The twisted structures are generated by starting from a 2H stacking ($\theta = 0^\circ$), which has the Mo (S) atom in the top layer directly above the S (Mo) in the bottom layer, and then rotating layers with the origin at an atom site\cite{Zhang2020}. The atomic structure of a moir\'e pattern of tb-{\mos} with $\theta=3.5^\circ$ is shown in Fig. \ref{fig:structre}(a), which contains 1626 atoms. In this paper, we mainly focus on plasmonic properties of tb-{\mos} with $\theta=3.5^\circ$ and $\theta=5.1^\circ$. The fully atomic relaxations are simulated via \red{the} Large-scale Atomic/Molecular Massively Parallel Simulator (LAMMPS)\cite{lammps1995computational} with the intralayer Stilliner-Weber  potential\cite{Jiang_2015} and the interlayer Lennard-Jones potential\cite{lammps-lj}. The relaxation effects on flat bands of tb-{\mos} are investigated in previous works\cite{tb-mos2018banddft,Zhan2020,flatband2021tb-tmd}. 
Here, we employ an accurate multi-orbital TB model to investigate the plasmons of tb-{\mos}. In this TB model, one unit cell of monolayer transition metal dichalcogenides comprises 11 orbitals, 5 d orbitals from one Mo atom, and 6 p orbitals from two S atoms\cite{Fang2015}. The total Hamiltonian of twisted bilayer {\mos} can be written as 
\begin{equation}
	\hat H = \hat H_1^{(1L)}+\hat H_2^{(1L)}+\hat H_{int}^{(2L)},
	\label{hal}
\end{equation}
where $ \hat H_{1(2)}^{(1L)}$ is the eleven-orbital single layer Hamiltonian, which contains the on-site energy, the hopping terms between orbitals of the same type at first-neighbor positions, and the hopping terms between orbitals of different type at first- and second-neighbor positions. The term $\hat H_{int}^{(2L)}$ is the interlayer interaction expressed as 
\begin{eqnarray}
	\hat H_{int}^{2L} = \displaystyle\sum_{p_i',\mathbf r_2,p_j,\mathbf r_1}\hat \phi_{2,p_i'}^\dagger(\mathbf r_2)t_{p_i',p_j}^{(LL)}(\mathbf r_2-\mathbf r_1)\hat \phi_{1,p_j}(\mathbf r_1) + \mathrm{H. c.},\nonumber\\
\end{eqnarray}
where $\hat \phi_{i,p_j}$ is the $p_j$ orbital basis of $i$-th monolayer. The interlayer hoppings in Slater-Koster (SK) relation are expressed with distance and angle as\cite{SK1954simplified}
\begin{equation}
	t_{p_i',p_j}^{(LL)}(\mathbf r) = (V_{pp,\sigma}(r)-V_{pp,\pi}(r))\frac{r_ir_j}{r^2}+V_{pp,\pi}(r)\delta_{i,j},
\end{equation}
where $r=|\mathbf r|$ and the distance-dependent SK parameter is
\begin{equation}
	V_{pp,b}=\nu_be^{[-(r/R_b)^{\eta_b}]},
	\label{inter}
\end{equation}
where $b=\sigma,\pi$, $\nu_b$, $R_b$ and $\eta_b$ are constant values taken from the Ref. \onlinecite{Fang2015}. In this paper, the interlayer interactions in twisted bilayer {\mos} are included in the TB Hamiltonian by adding hoppings between p orbitals of S atoms in the top and bottom layers with a distance smaller than 5 \AA~. The recent study shows that such a first-neighbor interlayer hopping approximation is appropriately enough\cite{isoband2020tbmos2}. When we relax the system, atoms move away from their equilibrium position in both in-plane and out-of-plane \red{directions}. As a consequence, we also need to change the intralayer hopping in Eq. (\ref{hal}). The intralayer hoppings in relaxed samples are modified with the form\cite{model2015intrahopping}
\begin{equation}
	t_{ij,\mu\nu}^{intra}(\mathbf r_{ij})=t_{ij,\mu\nu}^{intra}(\mathbf r_{ij}^0)\bigg(1-\Lambda_{ij,\mu\nu}\frac{|\mathbf r_{ij}-\mathbf r_{ij}^0|}{|\mathbf r_{ij}^0|}\bigg)
\end{equation}
where $t_{ij,\mu\nu}^{intra}$ is the intralayer hopping between the $\mu$ orbital of the $i$ atom and $\nu$ orbital of the $j$ atom, $\mathbf r_{ij}^0$ and $\mathbf r_{ij}$ are the distance between the $i$ and $j$ atoms in the equilibrium and relaxed cases, $\Lambda_{ij,\mu\nu}$ is the dimensionless bond-resolved local electron-phonon coupling. We assume that $\Lambda_{ij,\mu\nu}=3,4,5$ for the S-S $pp$, S-Mo $pd$, and Mo-Mo $dd$ hybridizations, respectively\cite{model2015intrahopping}. Note that a large Hamiltonian matrix describing a rigid or relaxed tb-{\mos} supercell will be generated. For example, the items in the Hamiltonian matrix of $3.5^\circ$ {\mos} are more than five thousand. Consequently, it is tough to diagonalize such a large matrix. Next, we will introduce \red{the} numerical methods of exploring plasmon properties in \red{the} hole-doped tb-{\mos}.

\subsection{Plasmon}

Polarization functions can be obtained from the Kubo formula\cite{kubo1957statistical}
\begin{equation}\label{kubo}
\begin{aligned}
\Pi_K(\mathbf{q},\omega)=& -\frac{2}{S}\int_{0}^{\infty}\mathrm dt\; e^{i\omega t}\mathrm{Im}\langle \varphi| n_F(H)e^{iHt}\\
&\times\rho(\mathbf{q})e^{-iHt}[1-n_F(H)]\rho(-\mathbf{q})|\varphi\rangle, 
\end{aligned}
\end{equation}
where $n_F(H)=\frac{1}{e^{\beta (H-\mu)}+1}$ is the Fermi-Dirac distribution operator, $\beta = \frac{1}{k_BT}$ being $T$ the temperature, $k_B$ the Boltzmann constant and $\mu$ the chemical potential. $\rho(\mathbf{q})=\sum_{i}c_i^{\dagger}c_i$exp$(i\mathbf{q}\cdot\mathbf{r}_i)$ is the density operator,  $\mathbf{r}_i$ is the position of $i$ orbital and $S$ is the area of a unit cell. As we mentioned before, each unit cell of tb-TMDs contains thousands of orbitals, which makes the diagonalization of the Hamiltonian very challenging. In this paper, we calculate the polarization function by combining the Kubo formula with a TBPM method. The TBPM is based on the numerical solution of time-dependent Schr\"{o}dinger equation and requires no diagonalization processes\cite{yuan2010tipsi}. By using the TBPM method, it is possible to obtain the electronic properties of large-scale systems, for instance, the \red{density of states (DOS)} of TBG with rotation angle $\theta$ down to $0.48^\circ$\cite{zhan2020large} and of dodecagonal graphene quasicrystal\cite{yu2019dodecagonal,dos_method}. The key idea in TBPM is to perform \red{an} average over initial states $|\varphi\rangle$, a random superposition of all basis states\cite{yuan2010tipsi, TBPM2000fast} 
\begin{equation}\label{random}
	|\varphi\rangle = \sum_{i}a_i|i\rangle,
\end{equation}
where ${|i\rangle}$ are all basis states in real space and $a_i$ are random complex numbers normalized as $\sum_{i}|a_i|^2 = 1$. By introducing the time evolution of two wave functions
\begin{eqnarray}
|\varphi_1(\mathbf{q},t)\rangle&&=e^{-iHt}[1-n_F(H)]\rho(-\mathbf{q})|\varphi\rangle,\\\nonumber
|\varphi_2(t)\rangle&&=e^{-iHt}n_F(H)|\varphi\rangle.
\end{eqnarray}
Then the real and imaginary parts of the dynamical polarization are
\begin{eqnarray}\label{tbpm}
\mathrm{Re}\Pi(\textbf{q}, \omega)&&=-\frac{2}{S}\int_{0}^{\infty}\mathrm dt\cos(\omega t)\mathrm{Im}\langle\varphi_2(t)|\rho(\mathbf{q})|\varphi_1(t)\rangle,\nonumber\\\\
\mathrm{Im}\Pi(\textbf{q}, \omega)&&=-\frac{2}{S}\int_{0}^{\infty}\mathrm dt\sin(\omega t)\mathrm{Im}\langle\varphi_2(t)|\rho(\mathbf{q})|\varphi_1(t)\rangle.\nonumber
\end{eqnarray}
The dynamical polarization function can be obtained from the Lindhard function as well\cite{linear2005quantum}
\begin{equation}\label{Lindhard}
	\begin{aligned}
		\Pi(\textbf{q}, \omega) = &\frac{g_s}{(2\pi)^2}\int_\mathrm{BZ}d^2\textbf{k}\sum_{l,l'}\frac{n_\mathrm{F}(E_{\mathbf{k'} l'}) - n_\mathrm{F}(E_{\mathbf{k}l})}
		{E_{\mathbf{k'} l' - E_{\mathbf{k}l}}-\omega-\mathrm{i}\delta}\\
		& \times |\langle \mathbf{k'} l'|\mathrm e^{\mathrm{i}\mathbf{q\cdot r}}|\mathbf{k}l \rangle |^2,
	\end{aligned}
\end{equation}
where $|\mathbf{k}l \rangle$ and $E_{\mathbf{k}l}$ are eigenstates and eigenvalues of the TB Hamiltonian in Eq. (\ref{hal}), respectively, with $\mathit{l}$ and $\mathit{l}'$ being band indices, $\mathbf{k'}$=$\mathbf{k}$+$\mathbf{q}$, $\delta \rightarrow 0^+$. Generally, the integral is taken over the whole first Brillouin zone (BZ) shown in Fig. \ref{fig:structre} (b).  It is convenient to analyze the contribution of band transitions to the polarization function as Eq. (\ref{Lindhard}) can be written as the sum of two parts 
\begin{equation}\label{intrainter}
	\Pi(\textbf{q}, \omega) = \Pi_{intra}(\textbf{q}, \omega) + \Pi_{inter}(\textbf{q}, \omega),
\end{equation}
where $\Pi_{intra}(\textbf{q}, \omega)$ and $\Pi_{inter}(\textbf{q}, \omega)$ denote intraband and interband contributions corresponding to  $\mathit{l} = \mathit{l}'$ and $\mathit{l} \neq \mathit{l}'$ in Eq. (\ref{Lindhard}), respectively. It is hard to sum over all bands obtained by diagonalizing TB Hamiltonian in Eq. (\ref{hal}) of a supercell that contains thousand of atoms. Therefore, we use the Eq. (\ref{tbpm}) to do full-band calculations. The validity of Eq. (\ref{tbpm}) has been verified by comparing the polarization function obtained from Eq. (\ref{tbpm}) and from a full-band calculation with Eq. (\ref{Lindhard})\cite{yuan2011kubo, Kuang2021}.

With the polarization function acquired from either Kubo formula in Eq. (\ref{tbpm}) or Lindhard function in Eq. (\ref{Lindhard}), the dielectric function that describes the electronic response to extrinsic electric perturbation, can be written within \red{the} random phase approximation (RPA) as 
\begin{equation}
\varepsilon(\textbf{q}, \omega) = 1- V(q)\Pi(\textbf{q}, \omega),
\label{dielectric}
\end{equation} 
in which $V(q)=2\pi e^2/{(\varepsilon_\mathrm{B}}{q})$ is the Fourier component of  two-dimensional Coulomb interaction, with $ \varepsilon_\mathrm{B}$ being the background dielectric constant. In our calculations, we set $\varepsilon_\mathrm{B}=3.03$ to represent the bulk dielectric constant of hexagonal boron nitride (hBN)\cite{hbn2018dielectric}. \red{The} electron energy loss (EL) function can be expressed as
\begin{equation}
S(\mathbf q, \omega)=-\mathrm {Im}(1/\varepsilon(\mathbf q, \omega)),
\label{loss}
\end{equation} 
which is an experimentally observable quantity to reflect the electronic response intensity. We can obtain \red{the} intraband EL function ($S_{intra}(\textbf{q}, \omega)$) or interband EL function ($S_{inter}(\textbf{q}, \omega)$)  by only taking $\Pi_{intra}(\textbf{q}, \omega)$ or $\Pi_{inter}(\textbf{q}, \omega)$ into account in Eq. (\ref{intrainter}). In this way, we can analyze intraband and interband transition contributions to \red{the} EL function by comparing $S_{intra}(\textbf{q}, \omega)$ and $S_{inter}(\textbf{q}, \omega)$ to $S(\textbf{q}, \omega)$, respectively. A plasmon mode with frequency $\omega_p$ and wave vector $\textbf{q}$ is well defined when a peak exists in \red{the} EL loss function at $\omega_p$.

\subsection{Density of states}
The density of states is calculated with TBPM as\cite{yuan2010tipsi, TBPM2000fast}
\begin{equation}\label{dos}
	D(E)=\lim\limits_{N \to \infty}\frac{1}{2\pi N}\displaystyle\sum_{p=1}^{N}\int_{-\infty}^{\infty}e^{iEt}\langle\varphi_p|e^{-iHt}|\varphi_p\rangle dt,
\end{equation}
where \textit N is the total number of initial states. In our calculations, the convergence of electronic properties can be guaranteed by utilizing a large enough system with more than 10 million atoms\cite{yuan2010tipsi}. 

\section{flat-band plasmons in relaxed tb-{\mos} with different twist angles}\label{relax3.5}
\begin{figure*}[htp]
\includegraphics[width=1\textwidth]{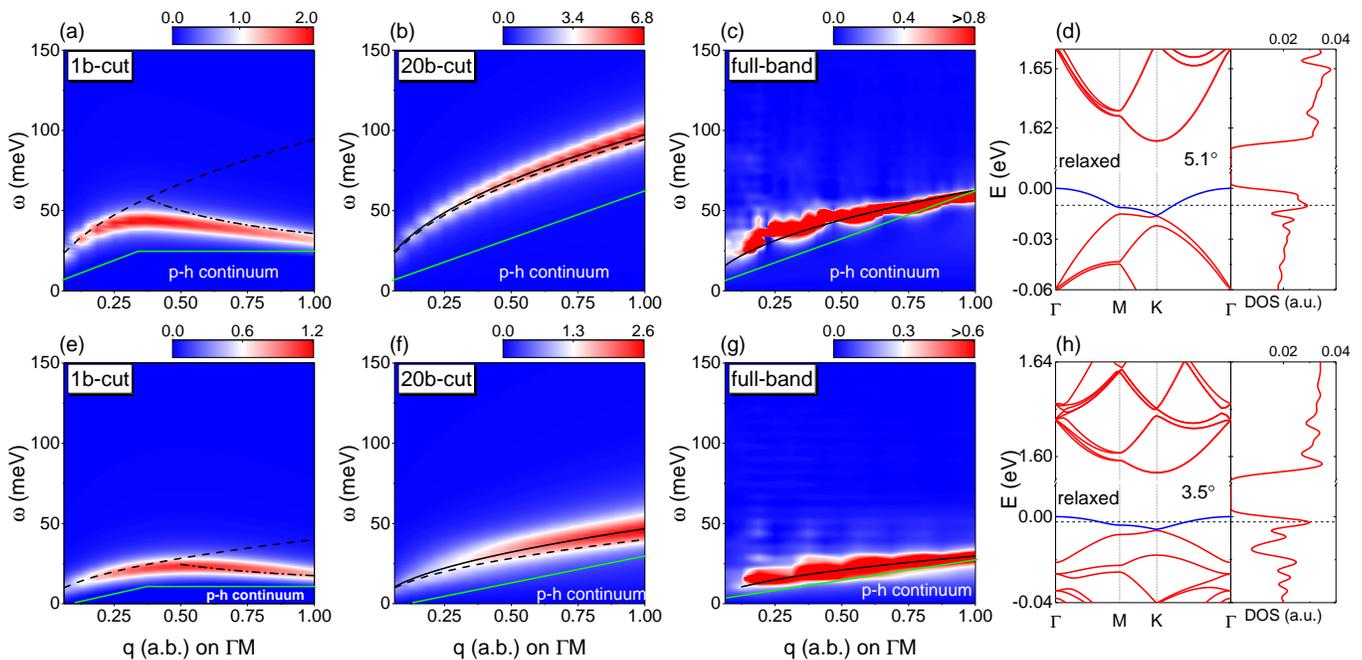}
\caption{EL function $S(\textbf{q},\omega)$ intensity plots of relaxed tb-{\mos} with (a)-(c) $\theta=5.1^\circ$ and (e)-(g) $\theta=3.5^\circ$  under different band cutoff calculations. Particle-hole (p-h) continuum region is marked with ``p-h continuum" and boundaries with green solid lines. Band structures and DOS for (d) $5.1^\circ$  and (h) $3.5^\circ$ are shown over a small energy range. The dashed lines crossing the flat bands in (d) and (h) denote chemical potentials, $\mu$= -10.0 meV and $\mu$= -2.4 meV, respectively. Here, the Fermi energy zero is set as the valence band maximum (VBM). In (a) and (e), the ``1b-cut" calculation means only the single doped flat-band (blue line) is included with $l=l'=1$ in Eq. (\ref{Lindhard}), while the ``20b-cut" calculation sums over 40 bands near zero energy (20 conduction bands and 20 valence bands) in (b) and (f). ``Full-band" calculations are performed via Eq. (\ref{tbpm}) to consider all the bands in (c) and (g). The maximum value for q (a.b.) = 1 represents the length from $\Gamma$ to M point denoted in Fig. \ref{fig:structre} and the minimum value of q is 0.0625$\gm$. Temperature is set to 1 K.}
	\label{fig: relaxplas}
\end{figure*}

In this section, we focus on flat-band plasmons in \red{the} relaxed hole-doped tb-{\mos}. A flat band (blue line) appears in the VBM at both $5.1^\circ$ and $3.5^\circ$, as shown in Figs. \ref{fig: relaxplas}(d) and \ref{fig: relaxplas}(h). The bandwidth $W$ of the flat band (\red{an} energy difference between the $\Gamma$ and K points of BZ) in $3.5^\circ$ ($W = 5.9$ meV) is much smaller than the one in $5.1^\circ$ ($W = 16.2$ meV). The density of states show high peaks, the van Hove singularities (VHS), at flat band energies. The doping levels with $\mu=-10.0$ meV and -2.4 meV in Figs. \ref{fig: relaxplas}(d) and (h) correspond to the near half filling of flat bands, respectively. In the EL function ($S(\textbf{q},\omega)$) spectra, particle-hole continuum ($\mathrm{Im}\Pi(\textbf{q}, \omega)<0$) regions are labeled by ``p-h continuum'' with boundaries ($\mathrm{Im}\Pi(\textbf{q}, \omega)=0$) illustrated by green solid lines (details in Appendix \ref{app-dyn}). The first and second rows in Fig. \ref{fig: relaxplas} show the results of tb-{\mos} with $5.1^\circ$ and $3.5^\circ$, respectively. The results in Figs. \ref{fig: relaxplas}(a)-(b) and \ref{fig: relaxplas}(e)-(f) are obtained from the Lindhard function in Eq. (\ref{Lindhard}). Full-band calculation results in Figs. \ref{fig: relaxplas}(c) and \ref{fig: relaxplas}(g) are performed via the Kubo formula in Eq. (\ref{tbpm}). The spectra with notation ``1b-cut" are calculated by only considering the single doped flat band, and the spectra with notation ``20b-cut''  are obtained by summing over 40 bands near zero energy (20 conduction bands (CBs) and 20 valence bands (VBs)) in Eq. (\ref{Lindhard}).

In the 1b-cut calculation, only intraband transitions with possible transition energies $\omega$ ($0< \omega < W$) are taken into account, whereas interband transitions between the doped flat band and other bands are neglected in Eq. (\ref{intrainter}). In this case, as shown in Figs. \ref{fig: relaxplas}(a) and \ref{fig: relaxplas}(e), the plasmons show quasi-flat dispersions, and are free from damping into electron-hole pairs as the plasmons locate above \red{the} p-h continuum region. Such unique dispersion can be well understood via a finite-bandwidth two-dimensional electron gas model (FBW-2DEG) (details in Appendix~\ref{app-Pi}). In the long wavelength limit $q < 0.25\gm$ and $q < 0.45\gm$ in Figs. \ref{fig: relaxplas}(a) and \ref{fig: relaxplas}(e), respectively, the plasmon dispersion can be well fitted with an ideal 2DEG model\cite{plas2020isolate}
\begin{equation}\label{2DEG}
		\omega_{pl} = \sqrt{\frac{2\pi n e^2q}{m\varepsilon_\mathrm{B}}},
\end{equation}
where n is the charge density related to a chemical potential $\mu$. The effective mass $m$ of the flat band at $\mu$ is obtained by fitting the band from $\Gamma$ to $M$ as a parabolic band. Then we obtain $m/m_e \approx -3.24$ at $\mu = -10.0$ meV and $m/m_e \approx -4.17$ at $\mu = -2.4$ meV. The dashed curves ($\omega_{pl} = a\sqrt{q}$) with \red{the} coefficients $a_{5.1}^{1b} = 92.1$ meV in Fig. \ref{fig: relaxplas}(a)-(b) and $a_{3.5}^{1b} = 33.4$ meV in Fig. \ref{fig: relaxplas}(e)-(f) are obtained via Eq. (\ref{2DEG}).

When $q > 0.25\gm$ and $q  > 0.45\gm$ in Figs. \ref{fig: relaxplas}(a) and \ref{fig: relaxplas}(e) plasmons deviate from $\sqrt{q}$ relation and show slightly negative dispersions. The reason is that the flat bands in Fig. \ref{fig: relaxplas} are not infinite parabolic bands but have finite bandwidths. The slightly negative dispersion can be well fitted by an analytical plasmon energy expression in the FBW-2DEG model\cite{plas2020isolate}
\begin{equation}\label{neg}
		\overline{\omega}_p =  \sqrt{\frac{\mu(2E_c -\mu)}{\text{exp}(q/q_{TF}) -1} + E_c^2},
\end{equation}
with $|E_c|$ as an effective finite bandwidth of the flat band and $q_{TF}$, the two-dimensional Thomas-Fermi vector gives as
\begin{equation}\label{TFvector}
	q_{TF} = \frac{2\pi e^2}{\varepsilon_\mathrm{B}} D(\mu),
\end{equation}
where $D(\mu)$ is the DOS value at $\mu$. The calculated $q_{TF}= 14.71$ nm$^{-1}$ with $\mu = -10.0$ meV at $5.1^\circ$ is very close to the value 14.77 nm$^{-1}$ with $\mu = -2.4$ meV at $3.5^\circ$. The two curves (dot-dashed lines) in Figs. \ref{fig: relaxplas}(a) and \ref{fig: relaxplas}(e) are obtained by setting $E_c = \mu = -10.0$ meV at $5.1^\circ$ and $E_c = E_M = -3.9$ meV at $3.5^\circ$($E_M$ is \red{the} flat-band energy at M point), respectively. Here, plasmon modes in \red{the} 1b-cut calculations are governed simply by intraband transitions inside the flat band, verifying that the single flat band guarantees undamped quasi-flat plasmons in the highly simplified one-band model. Moreover, the quasi-flat plasmon energy in Fig. \ref{fig: relaxplas}(e) is lower than that in Fig. \ref{fig: relaxplas}(a) due to the decrease of the flat-band width with reduced twist angles. 

 As seen from the band structure in Figs. \ref{fig: relaxplas}(d) and \ref{fig: relaxplas}(h), it is obvious that the doped flat bands are not completely separated from other VBs. In principle, both the transitions within flat bands and the effects of interband transitions on the flat-band plasmons should be considered. When 40 bands are considered in the polarization function (20b-cut calculation), plasmons with energy $\omega_p^{20b}$ exhibit $a^{20b}\sqrt{q}$ dispersion (black solid lines) in Figs. \ref{fig: relaxplas}(b) and \ref{fig: relaxplas}(f), and are away from the Landau damping regions. The coefficients $a_{5.1}^{20b}=98.3$ meV and $a_{3.5}^{20b}=38.4$ meV  slightly exceed those in the 2DEG mode \red{(dashed lines)}. Comparing the plasmon modes in Figs. \ref{fig: relaxplas}(b) and \ref{fig: relaxplas}(f) to those in Figs. \ref{fig: relaxplas}(a) and \ref{fig: relaxplas}(e), respectively, the effect of interband transitions on the doped flat-band plasmons is significant. That is, the inclusion of the interband transitions changes the quasi-flat dispersion of plasmon modes into $\sqrt{q}$ dispersion and dramatically enhances the energy of plasmons with \red{a} larger q. 

\begin{figure}[btp]
\includegraphics[width=0.5\textwidth]{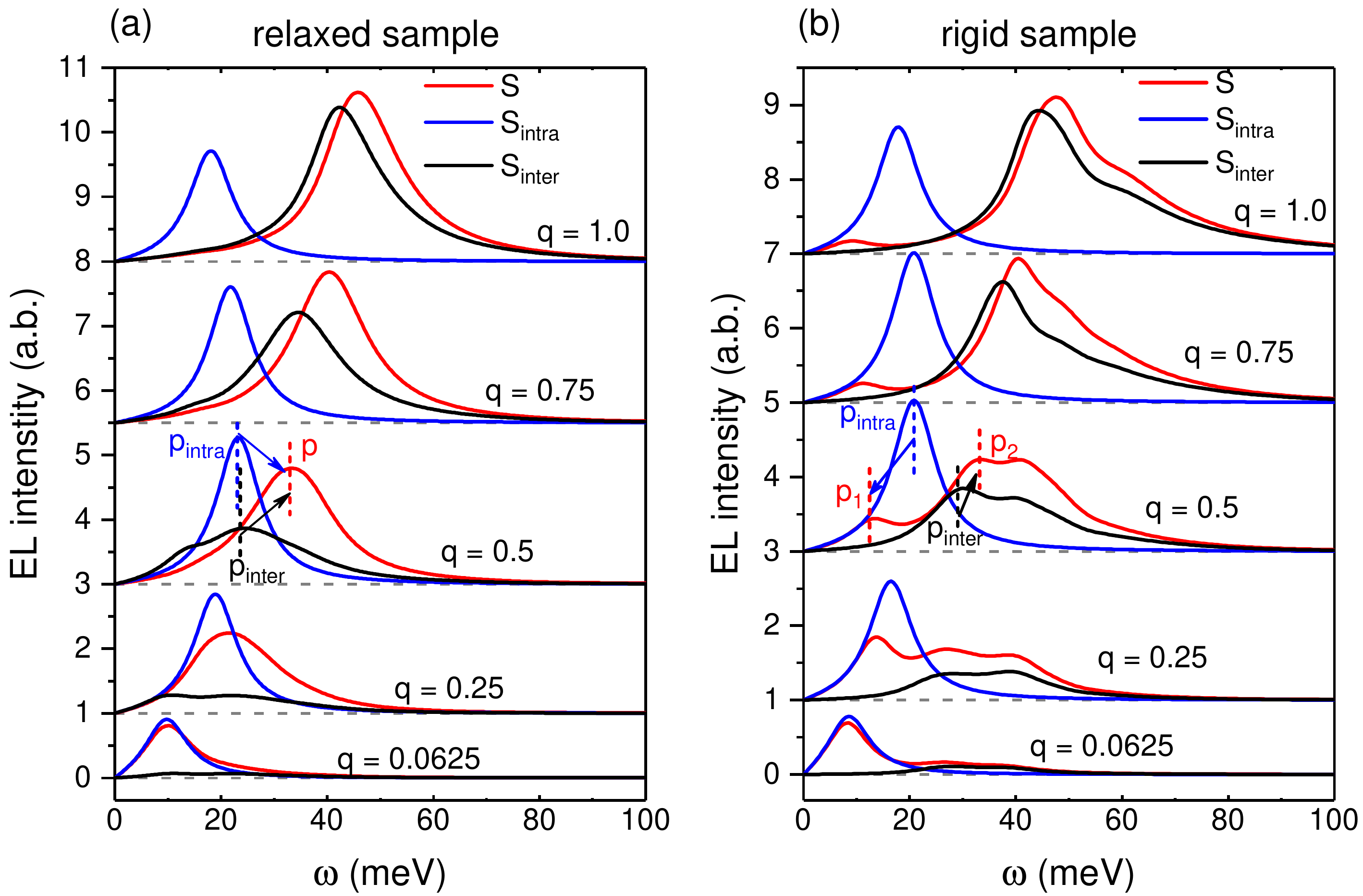}
\caption{EL functions $S$ (red solid lines), intraband EL functions $S_{intra}$ (blue solid lines) and interband EL functions $S_{inter}$ (black solid lines) at five sampled momenta q for (a) relaxed and (b) rigid tb-{\mos} with $\theta=3.5^\circ$ under 20b-cut calculation. $S$ are contributed by both intraband and interband transitions, while $S_{intra}$ and $S_{inter}$ are calculated by only taking intraband and interband transitions into account, respectively. Intraband and interband plasmon modes are marked by $p_{intra}$ (blue dashed lines) and $p_{inter}$ (black dashed lines), respectively. The notations $p$, $p_1$ and $p_2$ (red dashed line) represent the plasmon mode extracted from EL functions $S$. EL loss functions are shifted vertically for clarity and their zeros are denoted by gray dashed lines. } 
\label{fig:relaxloss}
\end{figure}

Previous works show that screening of high-energy interband transitions will decrease plasmon energies in monolayer and bilayer TMDs\cite{Louies2020universal,Nbse2012bandlfeffect,1l2lNbSe2013dftplas}. In order to figure out how \red{interband transitions will} modulate flat-band plasmons, in the 20b-cut calculation we compare EL functions $S$ (red lines) with intraband EL functions $S_{intra}$ (blue lines), and interband EL functions $S_{inter}$ (black lines) at sampled momenta $q$ for relaxed $3.5^\circ$ tb-{\mos} in Fig. \ref{fig:relaxloss}(a). The plasmon modes extracted from $S$, $S_{intra}$, and $S_{inter}$ are named as $p$, $p_{intra}$, and $p_{inter}$, respectively.
	For a small q = 0.0625$\gm$, the plasmon mode $p$ is overlapped with the intraband plasmon mode $p_{intra}$, which means that \red{the EL function $S$} is \red{solely dominated} by intraband transitions. For a large q =1.0$\gm$, \red{the} EL function $S$ has a similar shape with $S_{inter}$, implying that $p$ is mainly contributed by the interband plasmon mode $p_{inter}$. For q from 0.25$\gm$ to 0.75$\gm$, plasmon modes $p$ originate from both intraband and interband transitions and are affected by the interplay between $p_{intra}$ and $p_{inter}$. For example, when q = 0.5$\gm$, the non-zero parts of $S_{intra}$ and $S_{inter}$ are overlapped in an energy range ($0 < \omega < 50$ meV). The interplay of $p_{inter}$ and $p_{intra}$ yields \red{a mode with larger energy} (blue and black arrows) in Fig. \ref{fig:relaxloss}(a). This kind of interplay in relaxed tb-{\mos} is due to the fact that the flat band is not separated from other VBs, so \red{the} intraband transition energy can be overlapped with \red{the} interband transition energy in an energy range $0<\omega<W-|\mu|$.

\begin{figure*}[htp!]
	\includegraphics[width=1\textwidth]{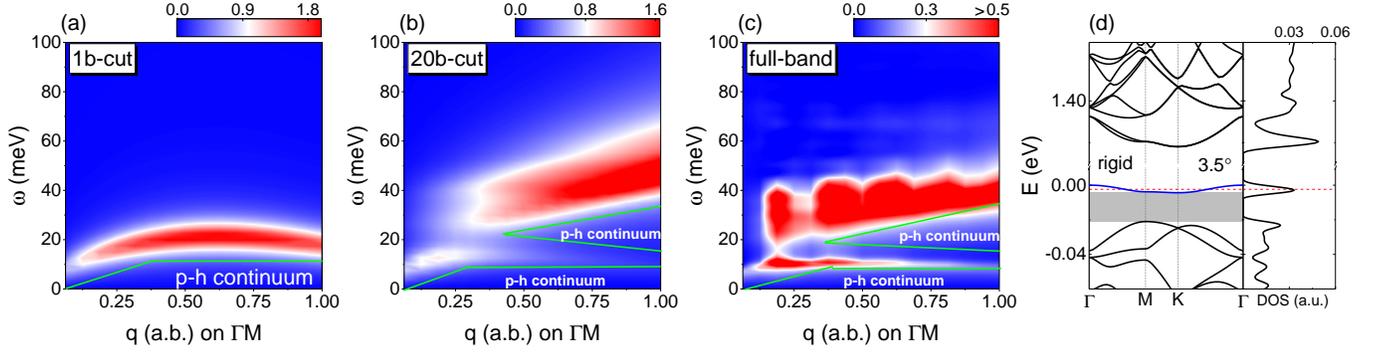}
	\caption{EL function intensity plots for rigid $3.5^\circ$ tb-{\mos} with chemical potential $\mu$= -2.0 meV (dashed line in (d)) and temperature T = 1 K, under (a) 1b-cut , (b) 20b-cut, and (c) full-band calculations. The p-h continuum regions with boundaries (green lines) are illustrated as well. (d) Band structure and DOS for rigid $3.5^\circ$ \mos. A band gap (shaded region) emerges between the doped flat VB (blue line) and other remote VBs.}
	\label{fig: rigidplas}
\end{figure*}

We further investigate how will interband transitions from much higher energy bands to the doped flat band affect the plasmonic properties. \red{As seen in Figs. \ref{fig: relaxplas}(c) and \ref{fig: relaxplas}(g),} \red{the} plasmon modes have lower energy with fitted $\sqrt{q}$ relation (solid lines) and tend to decay into p-h pairs at large momenta. The plasmon modes marked by black lines tend to be a linear dispersion with \red{a} larger q. Such tendency is caused by the screening effect of high-energy interband transitions on plasmons as aforementioned in previous works\cite{Louies2020universal,Nbse2012bandlfeffect,1l2lNbSe2013dftplas,pnasplas2019intrinsically}.  Next, we qualitatively explain these phenomena via an expression for plasmon energy\cite{pnasplas2019intrinsically}
\begin{equation}\label{BqAq}
	\omega_p ^2 \approx \frac{B(\textbf{q})}{1 + A(\textbf{q})},
\end{equation}
where $B(\textbf{q})$ contains the contribution of band transitions with \red{the} transition energy satisfying $|E_{\mathbf{k'} l'} - E_{\mathbf{k}l}| < \omega_p$, while $A(\textbf{q})$ is contributed by band transitions with relatively higher energies $|E_{\mathbf{k'} l'} - E_{\mathbf{k}l}| > \omega_p$ (detailed $A(\textbf{q})$ and $B(\textbf{q})$ in Appendix~\ref{app-Pi}). As shown in Fig. \ref{fig: relaxplas}, the plasmon modes in 20b-cut calculations have higher energies than that in the 1b-cut calculation. In the 20b-cut case, apart from a contribution of intraband transitions, the term $B(\textbf{q})$ has an extra contribution from the interband transitions with energies smaller than $\omega_p$, which results in an increment of the plasmon mode energy. Then, in the full-band calculation, the plasmon mode energy becomes smaller again because states from higher-energy bands (beyond the 40 bands) satisfy the condition $|E_{\mathbf{k'} l'} - E_{\mathbf{k}l}| > \omega_p$ and contribute to the term $A(\textbf{q})$. In full-band calculations, the plasmon energy becomes smaller when the twist angle decreases from $5.1^\circ$ \red{to} $3.5^\circ$. \red{The} twist angle effect on plasmon energy and flat bandwidth are similar (see Fig. \ref{fig:twistangle} in Appendix~\ref{app-angle}). Therefore, the flat-band plasmon can be a clue to detect the flat band.

In this part, we have analyzed the intraband and interband contributions to the plasmonic properties via \red{the three kinds of calculations} with different band cutoffs. The quasi-flat plasmon only appears in the one-band calculation, induced only by intraband transitions in the doped flat band. After that, if we consider more band effects, the plasmonic features are notably affected by \red{the} interband transitions. The effects of multi-band transitions on the flat-band plasmons in tb-{\mos} are different from that in TBG\cite{pnasplas2019intrinsically}. Here, \red{the} lower-energy quasi-flat plasmon dispersion in the simplified one-band calculation changes to higher-energy $\sqrt{q}$ \red{relation} in \red{both} multi-band and full-band calculations. However, for magic-angle TBG, the plasmon dispersion changes in a contrary way after considering more bands. That is, \red{the} classical plasmons with $\sqrt{q}$ relation in a simplified toy model (only including two flat bands), alter to \red{the} lower-energy quasi-flat plasmons obtained from a multi-band continuum model or full-band TB model\cite{pnasplas2019intrinsically,Kuang2021}. The different band cutoff effects on flat-band plasmons of TBG and tb-{\mos} could originate from the different features of the flat bands in the two twisted systems. The flat bands in relaxed TBG are entirely separated from other bands with gaps at least two times larger than the bandwidth \cite{Kuang2021}, while the flat band in relaxed tb-{\mos} only detaches from conduction bands above zero energy but touches its adjacent VB at K point, as shown in Figs. \ref{fig: relaxplas}(d) and \ref{fig: relaxplas}(h). So extra interband transitions in \red{the} multi-band calculation contribute to $A(\textbf{q})$  in magic-angle TBG\cite{pnasplas2019intrinsically} but to $B(\textbf{q})$ term in relaxed flat-band tb-{\mos}.

\section{flat-band plasmons in rigid tb-{\mos} with $\theta=3.5^\circ$}\label{rigid3.5}

We further study the effect of the lattice relaxation on plasmons in tb-{\mos}. For tb-{\mos} with $\theta = 3.5^\circ$ without relaxation (rigid tb-{\mos}), the flat band (blue line) with bandwidth $W=4.3$ meV is completely separated from other bands, as shown in Fig. \ref{fig: rigidplas}(d). The band gap $\Delta$ between the flat band and other VBs (\red{the} shaded region in Fig. \ref{fig: rigidplas} (d)) is 15.8 meV, three times larger than the bandwidth $W$. The plasmon spectra obtained via 1b-cut, 20b-cut, and full-band calculations are shown in Figs. \ref{fig: rigidplas}(a)-(c) with chemical potential $\mu$ = -2.0 meV (dashed line) near half filling of the flat band. In this case, \red{a} quasi-flat plasmon dispersion with the energy around 20 meV appears in the 1b-cut calculation. Interestingly, such \red{a plasmon dispersion with nearly constant energy} also emerges in both 20b-cut and full-band calculations with low energies even though it tends to vanish at \red{a} larger q. Besides, higher-energy interband plasmons also appear when $q> 0.25\gm$ in \red{both} 20b-cut and full-band calculations. When q near 0.5$\gm$, the two  plasmon dispersions (one from the intraband transitions and the other from interband transitions) are separated and can coexist in the spectra. In 20b-cut and full-band calculations, the p-h continuum regions are separated into two parts due to the presence of the band gap $\Delta$ (see Fig. \ref{fig:phc}(b) in Appendix~\ref{app-dyn}). The quasi-flat plasmon modes are low-damped as they reside at the gap between the two p-h continuum regions.

To gain insights into the distinct plasmon features in relaxed and rigid cases, we compare the contribution of band transitions to EL functions under 20b-cut calculation. In Fig. \ref{fig:relaxloss}(b), a signifcant difference is the existence of two plasmon branches \red{compared to Fig. \ref{fig:relaxloss} (a)}. The plasmon modes $p_1$ and $p_2$ correspond to the lower-energy quasi-flat and higher-energy plasmons in Fig. \ref{fig: rigidplas}(b), respectively, and $p_2$ is enhanced while $p_1$ is weakened with larger momenta. The two peaks $p_1$ and $p_2$ are contributed from intraband plasmon $p_{intra}$ and interband plasmon $p_{inter}$ (arrows in Fig. \ref{fig:relaxloss}(b)), respectively. For $q = 0.5\gm$, unlike the relaxed case where intraband and interband transitions \red{can be} superimposed in an energy range, the plasmons $p_{inter}$ and $p_{intra}$ always separate in the rigid case.

The profound explanation to the two plasmon modes is that due to the band gap $\Delta$ emerging in \red{the} rigid $3.5^\circ$ tb-{\mos}, interband transition energies $\omega> \Delta + W - |\mu|$ no longer overlap with intraband transition energies $W > \omega > 0$. As a result, $p_{intra}$ ($p_{inter}$) are softened (hardened) to $p_1$ ($p_2$) by the extra higher-energy interband transitions (lower-energy intraband transitions) contributing to $A(\textbf{q})$ ($B(\textbf{q})$), as shown with arrows in Fig. \ref{fig:relaxloss}(b). We can also find that \red{the} interband transitions play an important role in generating different plasmon features in relaxed and rigid tb-{\mos} from Fig. \ref{fig:relaxloss}, and \red{the} interband plasmons $p_{inter}$ gradually dominate plasmons with larger momenta, which could owe to the enhancement of \red{the} interband coherence factor in Eq. (\ref{Lindhard}) (see Fig. \ref{fig:corre} in Appendix~\ref{app-corre}).

In brief, the flat band can lead to quasi-flat and low-damped plasmons in the rigid sample. The intraband plasmons can coexist with the higher-energy interband plasmons at some momenta in \red{both} multi-band and full-band calculations. 
The presence of the band gap $\Delta$ ensures that \red{the} two plasmon branches appear \red{simultaneously} in \red{the} EL spectra. On the contrary, in the relaxed sample, due to the absence of the band gap between the flat band and its adjacent VBs, interband transitions start to contribute in a very tiny energy. As a result, the single plasmon mode has both contributions from the interband and intraband transitions. However, we detect a quasi-flat plasmon in relaxed tb-{\mos} with \red{an} angle smaller than $3.5^\circ$ (see Fig. \ref{fig:1.6degree}(a) in Appendix~\ref{app-angle}), at which a band gap $\Delta$ also appears. As a conclusion, the separation of the flat band from other bands in tb-{\mos} plays a crucial role in the exploration of quasi-flat and low-damped plasmons since the band gap affects the interband contribution to plasmons. 

\section{Band cutoff and doping effects}
\label{sec-bandcutmu}
\begin{figure}[htp!]
	\includegraphics[width=0.5\textwidth]{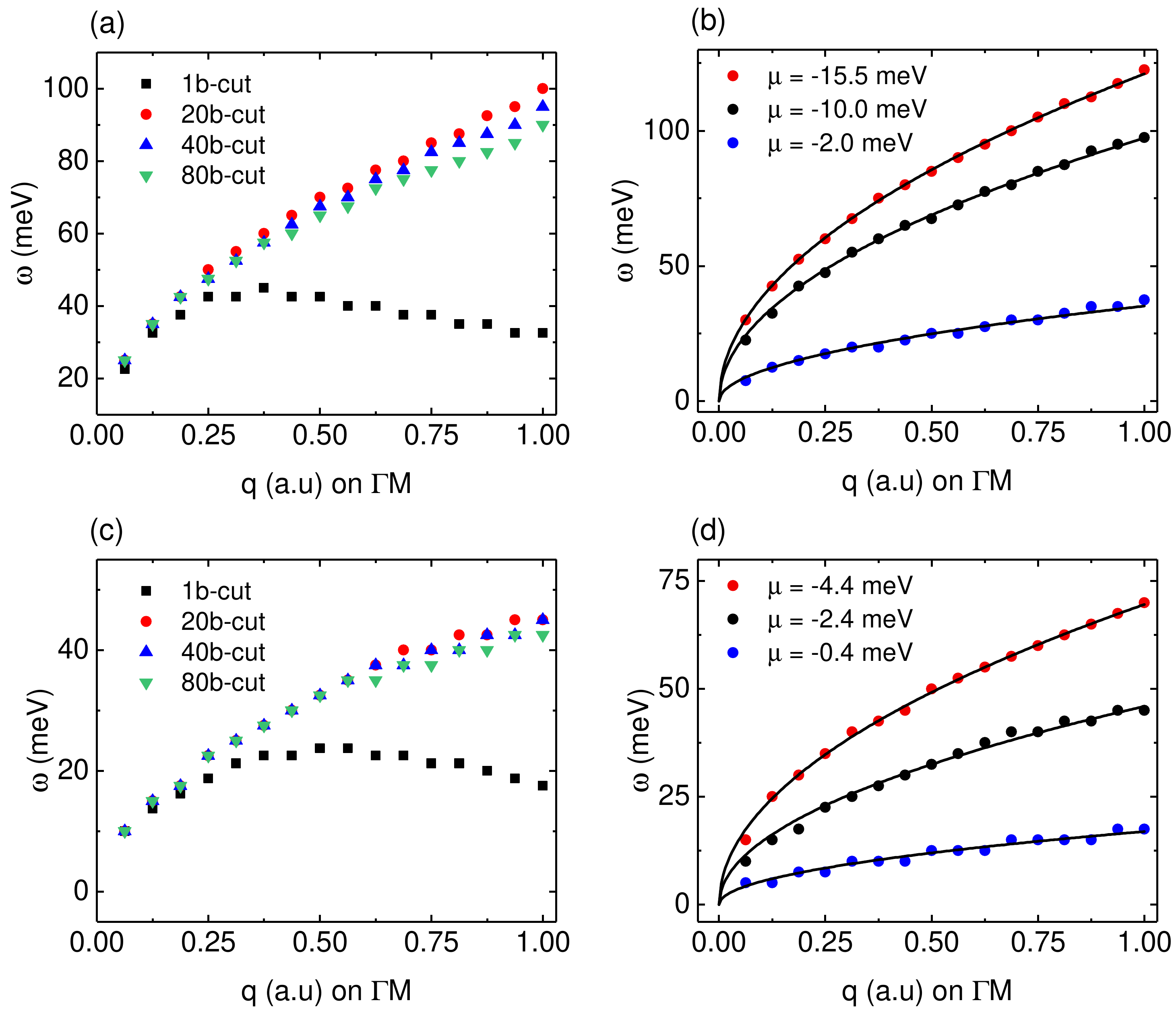}
	\caption{Band cutoff effect on plasmon energy versus q for relaxed tb-{\mos} (a) with $\theta=5.1^\circ$ and $\mu$ = -10 meV and (c) $\theta=3.5^\circ$ and $\mu$ = -2.4 meV. The doping effect on plasmon energy versus q for relaxed tb-{\mos} with (b) $5.1^\circ$ and (d) $3.5^\circ$ under the 20b-cut calculation.
		1b-cut (black square), 20b-cut (red circle), 40b-cut (blue triangle) and 80b-cut (green triangle) in (a) and (c) stand for the 1 band (the doped flat 
		VB), 40 bands (20 CBs above and 20 VBs below zero energy), 80 bands (40 CBs above and 40 VBs below zero energy), and 160 bands (80 CBs above and 80 VBs below zero energy) cutoff calculations via Eq. (\ref{Lindhard}) with $l$ and $l'$ summing over 1 band, 40 bands, 80 bands and 160 bands, respectively. Plasmon energy changes with three different chemical potentials in (b) and (d) with 20b-cut calculations via Eq. (\ref{Lindhard}).} 
	\label{fig:mucutcom}
\end{figure}
In this part, we move forward to investigating the band cutoffs and doping effects on plasmons in \red{the} relaxed tb-{\mos} with $5.1^\circ$ and $3.5^\circ$ in Fig. \ref{fig:mucutcom}. First, we compare the plasmons calculated with different band cutoffs in Fig. \ref{fig:mucutcom}(a) for $5.1^\circ$ and Fig. \ref{fig:mucutcom}(c) for $3.5^\circ$. The plasmon energy significantly increases from 1b-cut calculation (black squares) to 20b-cut calculation (red dots) with \red{a momentum q} getting larger. Then, after taking more bands (blue and green triangles) into account, the plasmon energy decreases when  $q > 0.25\gm$ and $q > 0.56\gm$ in Figs. \ref{fig:mucutcom} (a) and \ref{fig:mucutcom} (c), respectively. In 40b-cut and 80b-cut calculations, other higher-energy interband transitions contribute to $A(\textbf{q})$, and decrease the plasmon energy. 
For $q<0.25\gm$ in $5.1^\circ$ and $q < 0.56\gm$ in $3.5^\circ$, the plasmon energy converges even in the 20b-cut calculation. Therefore, only for limited small wave numbers, it is accurate enough to model the flat-band plasmon with an appropriate band cutoff calculation. This also implies that if plasmon in relaxed tb-{\mos} is studied via a low-energy continuum model \cite{tb-mos2021kpmodel,tb-mos2021contituummodel,tb-tmd2021conti128band}, the plasmon energy will be overestimated at a larger twist angle and momentum. The low-energy continuum model only accurately describes a finite number of bands near \red{the} Fermi energy, which neglects the effects of interband polarization of higher electron bands. Such \red{an} overestimation of \red{the} plasmon energy could affect the prediction of plasmon-mediated superconductivity\cite{prs2020superconductivity, lewandowski2020pairing}.

\begin{figure}[t!]
	\includegraphics[width=0.5\textwidth]{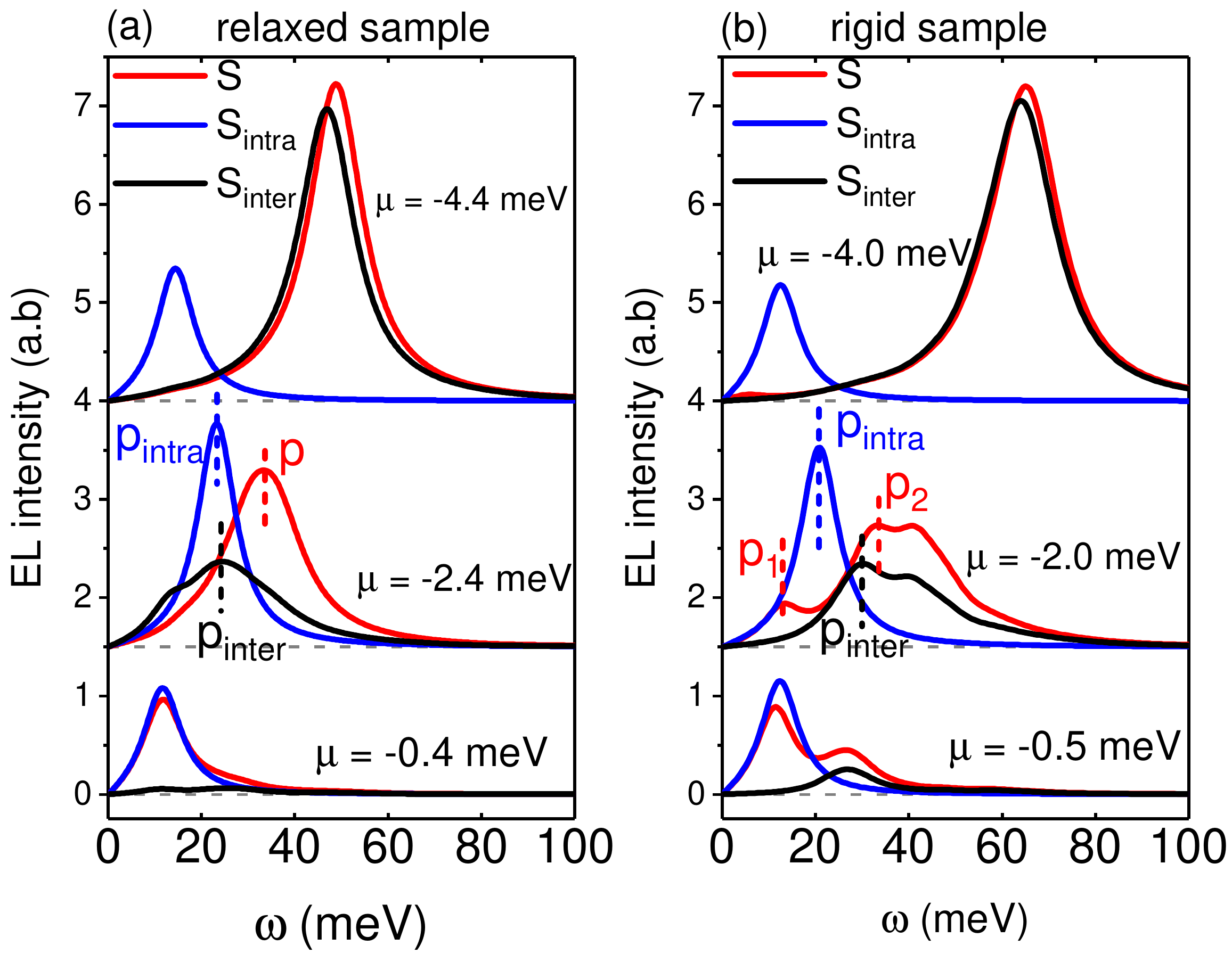}
	\caption{EL functions $S$ (red solid lines), intraband EL functions $S_{intra}$ (blue solid lines), and intrerband EL functions $S_{inter}$ (black solid lines) at different chemical potentials $\mu$, for (a) relaxed and (b) rigid tb-{\mos} with $3.5^\circ$ under 20b-cut calculations. Intraband and interband plasmon modes are marked by $p_{intra}$ (blue dashed lines) and $p_{inter}$ (black dashed lines), respectively. The notations $p$, $p_1$ and $p_2$ (red dashed line) represent the plasmon mode extracted from EL functions $S$. EL loss functions are shifted vertically for clarity and their zeros are denoted by gray dashed lines.} 
	\label{fig:muloss}
\end{figure}

Next, we show that modulating chemical potential $\mu$ is another way to change the contribution of interband transitions to plasmons dramatically. In \red{the} relaxed tb-{\mos} with $5.1^\circ$ in Fig. \ref{fig:mucutcom}(b) and \red{with} $3.5^\circ$ in Fig. \ref{fig:mucutcom}(d), plasmon energies $\omega_p^{20b}$ (circles) with different $\mu$ are obtained via 20b-cut calculations. The results are also fitted with $\sqrt{q}$ curves (solid black lines).
%
On the one hand, decreasing the magnitude of chemical potential $\mu$ leads to smaller plasmon energy. The plasmon energy tends to be constant at large momenta when \red{the} doping level closes to 0, as shown in Fig. \ref{fig:mucutcom}(d) with blue dots. With \red{a} larger hole doping introduced, interband transitions are enhanced via modulating the Fermi-Dirac factor in Eq. (\ref{Lindhard}) (see Fig. \ref{fig:fermi} in Appendix~\ref{app-corre}), which results in larger plasmon energies at higher hole-doping levels (black and red circles) in Figs. \ref{fig:mucutcom}(b) and \ref{fig:mucutcom}(d). This can be further verified by investigating how \red{will} intraband plasmon $p_{intra}$ and interband plasmon $p_{inter}$ contribute to plasmon $p$ in EL functions at different doping levels with a sampled $q = 0.5\gm$, as seen in Fig. \ref{fig:muloss}(a). The interband plasmon modes $p_{inter}$ monotonously increase with $|\mu|$, while the intraband plamon mode $p_{intra}$ has the maximum energy near half filling of the flat band. Furthermore, as shown in Fig. \ref{fig:muloss}(a), the plasmon mode $p$ is almost completely contributed by $p_{intra}$ at $\mu$ = -0.4 meV, then generated by both of $p_{intra}$ and $p_{inter}$ when $\mu$ = -2.4 meV, and mainly attributed by $p_{inter}$ at $\mu$ = -4.4 meV with the flat band almost fully hole-doped. In fact, \red{the} low-energy intraband plasmons $p_{intra}$ dominate \red{the} plamsons $p$ for most of momenta (except for q near $\gm$) at $\mu$ = -0.4 meV, whereas \red{the} plasmons $p$ are mainly contributed by higher-energy interband plasmons $p_{inter}$ for all of q (even for q near 0) at $\mu$ = -4.4 meV (see Fig. \ref{fig:relaxmu5q} in Appendix~\ref{app-lossqu}). Therefore, we can conclude that \red{the} stronger and higher-energy interband plasmons at larger hole-doping levels play key roles in enhancing the plasmon energies in Figs. \ref{fig:mucutcom}(b) and \ref{fig:mucutcom}(d). There is still only one plasmon peak appearing in EL functions $S$ by tuning hole-doping levels. As a result, the quasi-flat plasmon mode is not observed in high and low hole-doping levels in relaxed tb-{\mos} with $3.5^\circ$. Thus, the separation of \red{the} flat band from other bands in tb-{\mos} is still the key to explore quasi-flat and low-damped plasmon modes.

\begin{figure}[t!]
	\includegraphics[width=0.5\textwidth]{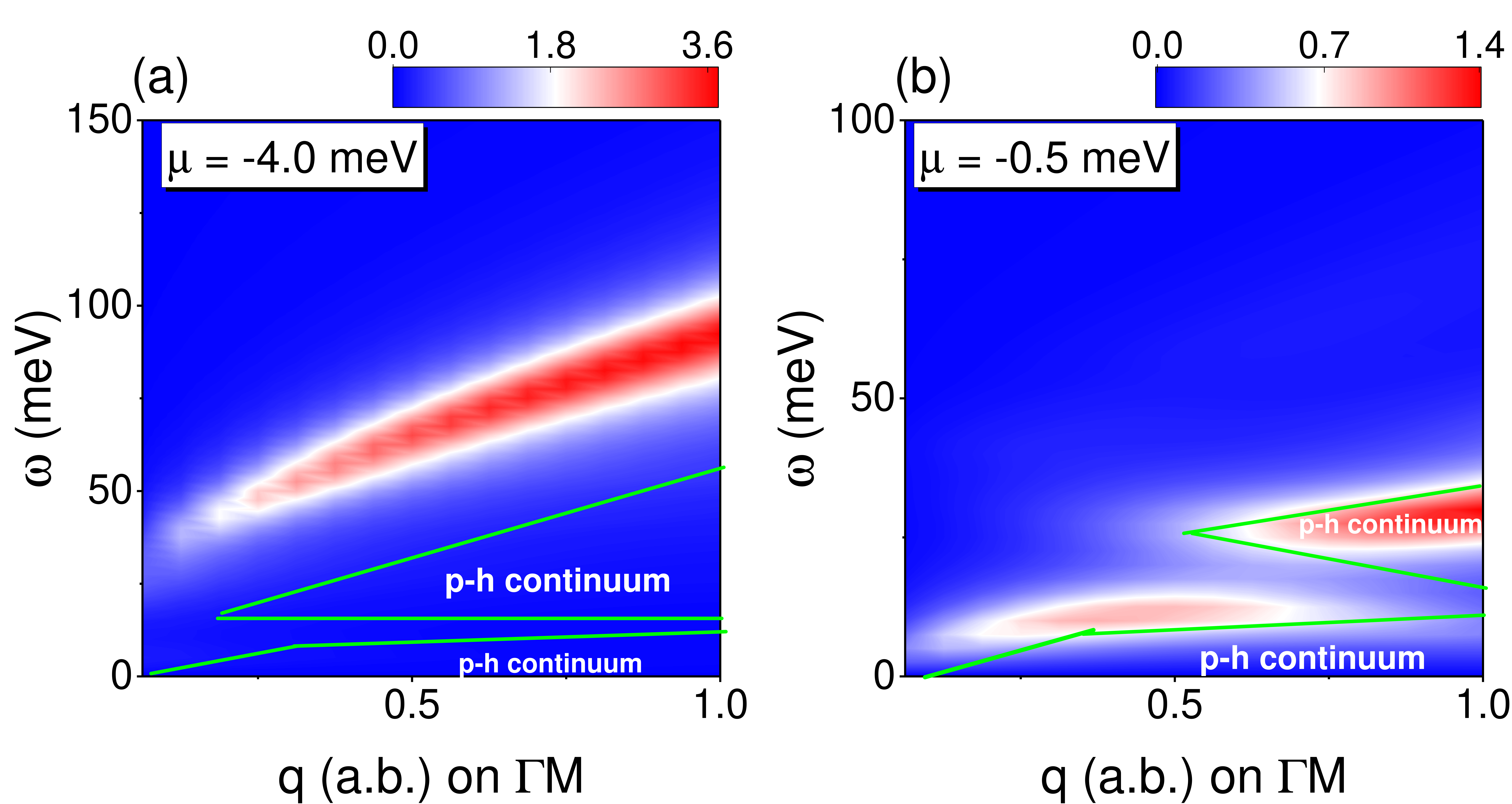}
	\caption{EL function intensity spectra for rigid $3.5^\circ$ tb-{\mos} with chemical potential (a) $\mu$= -4.0 and (b) $\mu$ = -0.5 meV (b) in 20b-cut calculations. The p-h continuum region with boundaries (green lines) are also illustrated.} 
	\label{fig:rigidmuplas}
\end{figure}

Based on the fact that plasmon dispersion are obviously altered by $\mu$ in relaxed cases, we turn to study how \red{the} quasi-flat plasmons appearing in rigid case are influenced by chemical potentials. When the isolated flat band is slightly doped with $\mu$ = -0.5 meV, both quasi-flat plasmons and higher-energy interband plasmons still exist in Fig. \ref{fig:rigidmuplas}(b). The quasi-flat plasmons are still low-damped, whereas the interband plasmons are overdamped. Once tuning $\mu$ to -4.0 meV with the flat band nearly full filled, only one undamped plasmon dispersion appears in Fig. \ref{fig:rigidmuplas}(a). To unveil how quasi-flat plasmons are affected by doping levels, we study \red{the} intraband and interband contribution to plasmons at the three hole-doping levels with a fixed q = 0.5$|\Gamma M|$ in Fig. \ref{fig:muloss}(b). First, \red{the} interband plasmons $p_{inter}$ are enhanced with larger hole-doping levels. From $\mu$ = -0.5 to -2.0 meV, the enhancement of $p_{inter}$ at $\mu$ = -2.0 meV weakens $p_1$, despite of stronger $p_{intra}$ with larger energy compared to $\mu$ = -0.5 meV. The quasi-flat plasmon spectra weights in Fig. \ref{fig: rigidplas}(b) thus become weaker comparing to Fig. \ref{fig:rigidmuplas}(b). Besides, such \red{an} enhanced $p_{inter}$ with higher energy also causes the disappearance of the quasi-flat plasmons $p_1$ and the only existence of higher-energy plasmons at $\mu$ = -4.0 meV in Fig. \ref{fig:rigidmuplas}(a). In fact, \red{the} plasmon mode $p_1$ arising from $p_{intra}$ is visible only for q = 0.0625$\gm$ (see Fig. \ref{fig:rigidmu5q}(b) in Appendix~\ref{app-lossqu}). The emergence of $p_1$ is because much weaker higher-energy interband transitions at the smallest q does not completely quench intraband plasmon $p_{intra}$ via the term $A(\textbf{q})$ in Eq. (\ref{BqAq}), as discussed in Fig. \ref{fig:relaxloss}(b). We can also see that the single plasmon dispersion with $\mu$ = -4.0 meV is almost completely contributed by \red{the} interband plasmons $p_{inter}$ except for q = 0.0625$\gm$ (shown in Fig. \ref{fig:rigidmu5q}(b) in Appendix~\ref{app-lossqu}). Besides, the relatively weaker and higher-energy interband plasmons $p_{inter}$ at $\mu$ = -0.5 meV (see Fig. \ref{fig:rigidmu5q}(a) in Appendix~\ref{app-lossqu}) ensure that the intraband and interband plasmon modes coexist with $q > 0.5\gm$  in Fig. \ref{fig:rigidmuplas}(b). The chemical potential is essential for observing the coexistence of the two plasmon modes in the rigid case.

All in all, plasmonic properties in relaxed and rigid tb-{\mos} can be notably affected by interband transitions at different hole-doping levels. The quasi-flat plasmons can be killed with a large hole-doping level since \red{the} intraband plasmons are strongly weakened by the enhanced higher-energy interband transitions at 
\red{a} larger $|\mu|$, which simultaneously contribute to \red{the} stronger and higher-energy interband plasmons. This also implies that a slighter hole-doping level is more conducive to observing the quasi-flat plasmons in rigid tb-{\mos}. Besides, the isolated flat band can not always promise the existence of quasi-flat plasmons because of the doping effect on interband contributions.

\section{summary and discussion}
In summary, we investigated flat-band plasmons in \red{the} hole-doped tb-{\mos} with and without \red{lattice} relaxations, and analyzed the intraband and interband contribution to plasmons. Different band cutoffs are considered in the polarization function to tune interband transitions between the single flat band and other bands. In relaxed cases, the flat band is not separated from other valence bands so that the interband and intraband transitions can interfere with each other in the low-energy range. When interband transitions are introduced in multi-band calculations, the quasi-flat plasmons emerging in the one-band calculation are transformed into classical plasmons of 2DEG with $\sqrt{q}$ dispersion. The full-band calculation, including higher-energy interband transitions, decreases plasmonic energies of \red{the} classical plasmons observed in \red{the} multi-band calculation. We also compared plasmons in \red{the} relaxed tb-{\mos} with $5.1^\circ$  to those with $3.5^\circ$. The plasmon energy becomes smaller when the twist angle decreases since a smaller angle gives rise to a flatter band. In \red{the} rigid tb-{\mos} with $3.5^\circ$, the flat band is separated from valence bands with a gap three times larger than its bandwidth. As a consequence, \red{the} interband and intraband transitions occur in different energy ranges. We observe two plasmon branches in \red{the} rigid tb-{\mos}. One is a lower-energy quasi-flat plasmon (intraband plasmon), and the other is a higher-energy plasmon (interband plasmon). Moreover, the quasi-flat plasmons can be observed in one-band, multi-band, and full-band calculations. For other tb-TMDs, for example, twisted bilayer MoSe$_2$, twisted bilayer WS$_2$ and twisted bilayer WSe$_2$, such a band gap also disappears in relaxed cases\cite{flatband2021tb-tmd}. Therefore, similar plasmon properties could be observed in these tb-TMDs.

Besides, different band cutoffs in multi-band calculations can change the plasmon energy at \red{a} large q in \red{the} relaxed tb-{\mos}. Tuning hole-doping levels can notably
	change plasmon energy in relaxed tb-{\mos} and affect
	the coexistence of two plasmon branches in rigid tb-{\mos}, as interband contributions to plasmon can be significantly tuned by \red{the} doping of the flat band. Plasmons are gradually dominated by \red{the} enhanced interband transitions with more holes filled in the flat band. When the flat band is almost full-filled with holes, only one interband plasmon dispersion is observed in \red{both} rigid and relaxed cases, and the quasi-flat plasmons disappear in \red{the} rigid tb-{\mos}. In the future, flat band systems remain potential platforms to explore undamped, low-energy dispersionless plasmons and their applications like plasmonic superconductivity. Based on the analysis of band cutoff calculation, we also need to think about the validity of low-energy models for studying flat-band plasmons in twisted two-dimensional semiconductors, especially when the interband transitions play a dominant role. 

\begin{acknowledgments}
We thank Francisco Guinea for his valuable discussions. This work was supported by the National Natural Science Foundation of China (Grants No. 12174291 and No. 12047543). S.Y. acknowledges funding from the National Key R\&D Program of China (Grant No. 2018YFA0305800). X.K. acknowledges the financial support from China Scholarship Council (CSC). Numerical calculations presented in this paper have been performed on the supercomputing system in the Supercomputing Center of Wuhan University.
\end{acknowledgments}

\appendix
\section{Twist angle effect in relaxed tb-{\mos}}
\label{app-angle}
\begin{figure}[htb!]
	\includegraphics[width=0.5\textwidth]{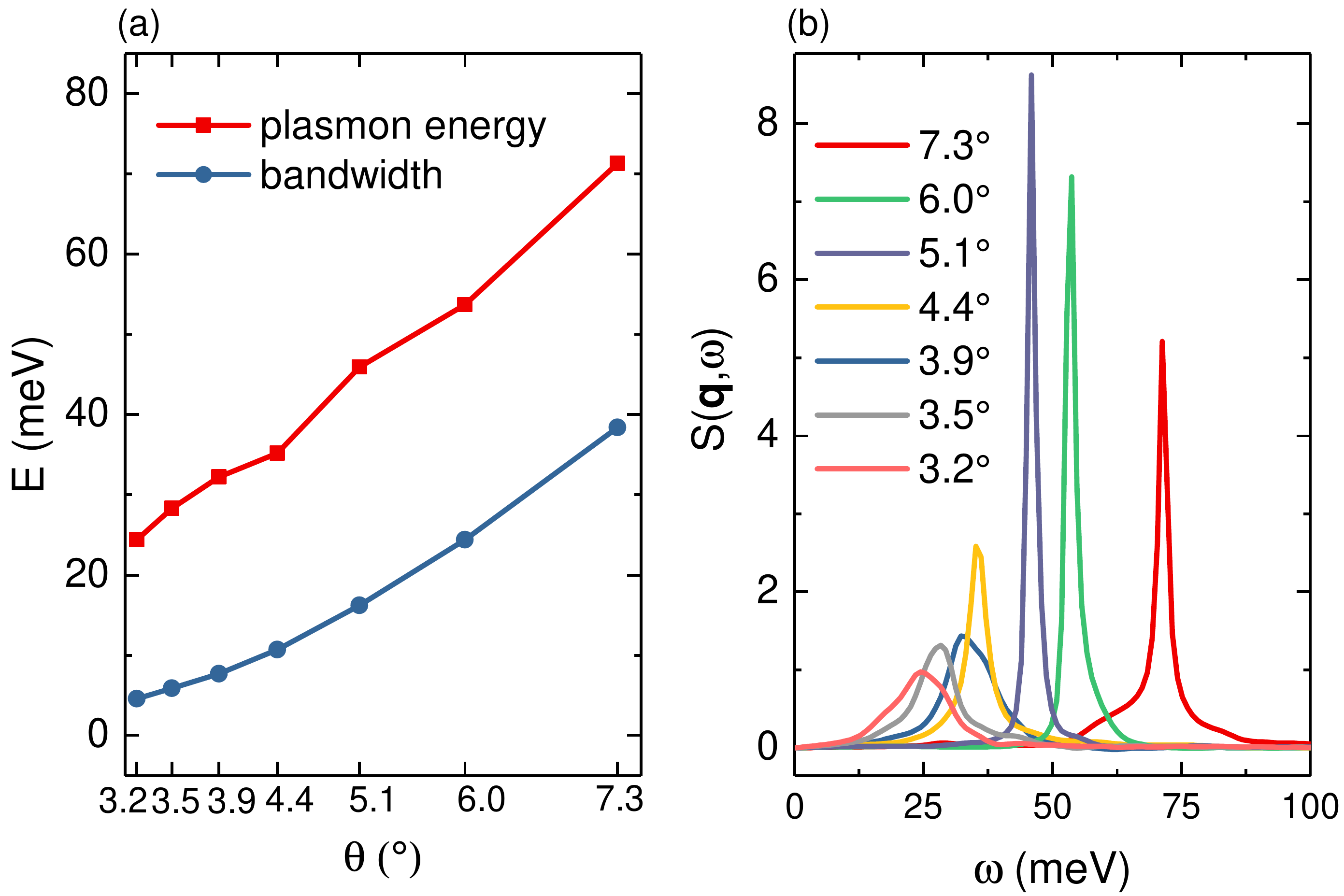}
	\caption{(a) Plasmon energy and bandwidth of the flat band versus twist angles $\theta$ from $3.2^\circ$ to $7.3^\circ$ in relaxed tb-{\mos}. (b) EL functions of relaxed tb-{\mos} at different angles under full-band calculation with $q = 0.5|\Gamma M|$ and T =1K. The chemical potential makes half filling of the flat band at each twist angle. Plasmon energies in (a) are extracted from the peaks of EL function in (b).} 
	\label{fig:twistangle}
\end{figure}

\begin{figure}[t!]
	\includegraphics[width=0.5\textwidth]{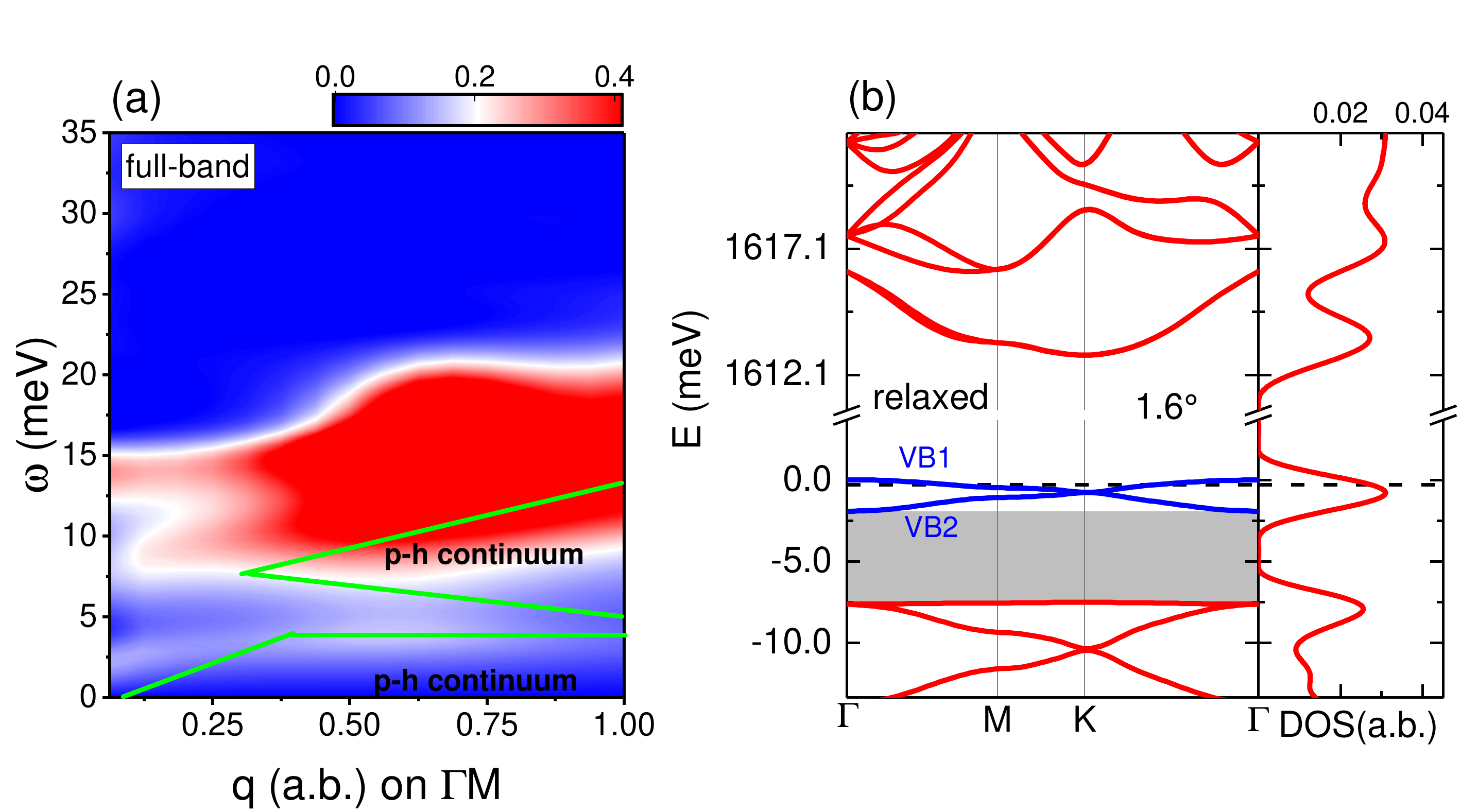}
	\caption{(a) EL function intensity plot and p-h continuum region with boundaries (green lines) for relaxed tb-{\mos} with $1.6^\circ$ by adopting full-band calculation. (b) Band structure and DOS for $1.6^\circ$ relaxed tb-{\mos}. Two flat bands, VB1 and VB2 (blue lines), are separated from remote VBs with a gap (shaded region). The dashed line denotes the chemical potential $\mu$ = -0.3 meV in the calculation.} 
	\label{fig:1.6degree}
\end{figure}

The twist angle effect on plasmonic properties has been discussed by comparing plasmons at $5.1^\circ$ and $3.5^\circ$. To further unveil the relation between plasmon energy and twist angle, we obtain the plasmon energies in Fig. \ref{fig:twistangle}(a) under \red{the} full-band calculation for angles ranging from $3.2^\circ$ to $7.3^\circ$. At each angle, the flat band does not separate from other VBs as the one in $3.5^\circ$ and $5.1^\circ$. Only one plasmon mode appears in the EL functions at these angles, as seen in Fig. \ref{fig:twistangle}(b). The energy of the plasmon mode with \red{a} fixed q = 0.5$|\Gamma M|$ decreases as the twist angle reduces. In an experiment, the detected plasmon energies at different angles could thus reflect the distinct bandwidth of tb-{\mos}, as discussed in TBG\cite{Kuang2021}. However, the single plasmon mode transforms into two separated plasmon modes when the flat bands are separated from the valence band with a gap $\Delta$ as in a rigid case \cite{flatband2021tb-tmd}. For example, there are two isolated flat bands (VB1 and VB2) near zero energy in \red{the} relaxed tb-{\mos} with $\theta = 1.6^\circ$ in Fig. \ref{fig:1.6degree}(b). We obtain the plasmon spectrum under \red{the} full-band calculation in this case after making the flat band VB1 near half filling, with the doping level $\mu$ = -0.3 meV (the dashed line Fig. \ref{fig:1.6degree}(b)). There are two kinds of plasmon modes in Fig. \ref{fig:1.6degree}(a); one is the \red{quasi-flat} plasmon with the energy around 3 meV, which is contributed by the band transitions in the two flat bands, while another is \red{the} interband plasmon arising from the interband transitions between the doped flat band and remote valence bands. Here, the plasmon spectra feature in \red{the} relaxed tb-{\mos} with $\theta = 1.6^\circ$ is similar to rigid  tb-{\mos} at $\theta = 3.5^\circ$ in Fig. \ref{fig: rigidplas}(c), but the flat-band plasmon modes are damped for entering the p-h continuum for $q > 0.25|\Gamma M|$. 

\section{Particle-hole continuum spectra in tb-{\mos} with $\theta=3.5^\circ$}
\label{app-dyn}
\begin{figure}[h!]
	\includegraphics[width=0.5\textwidth]{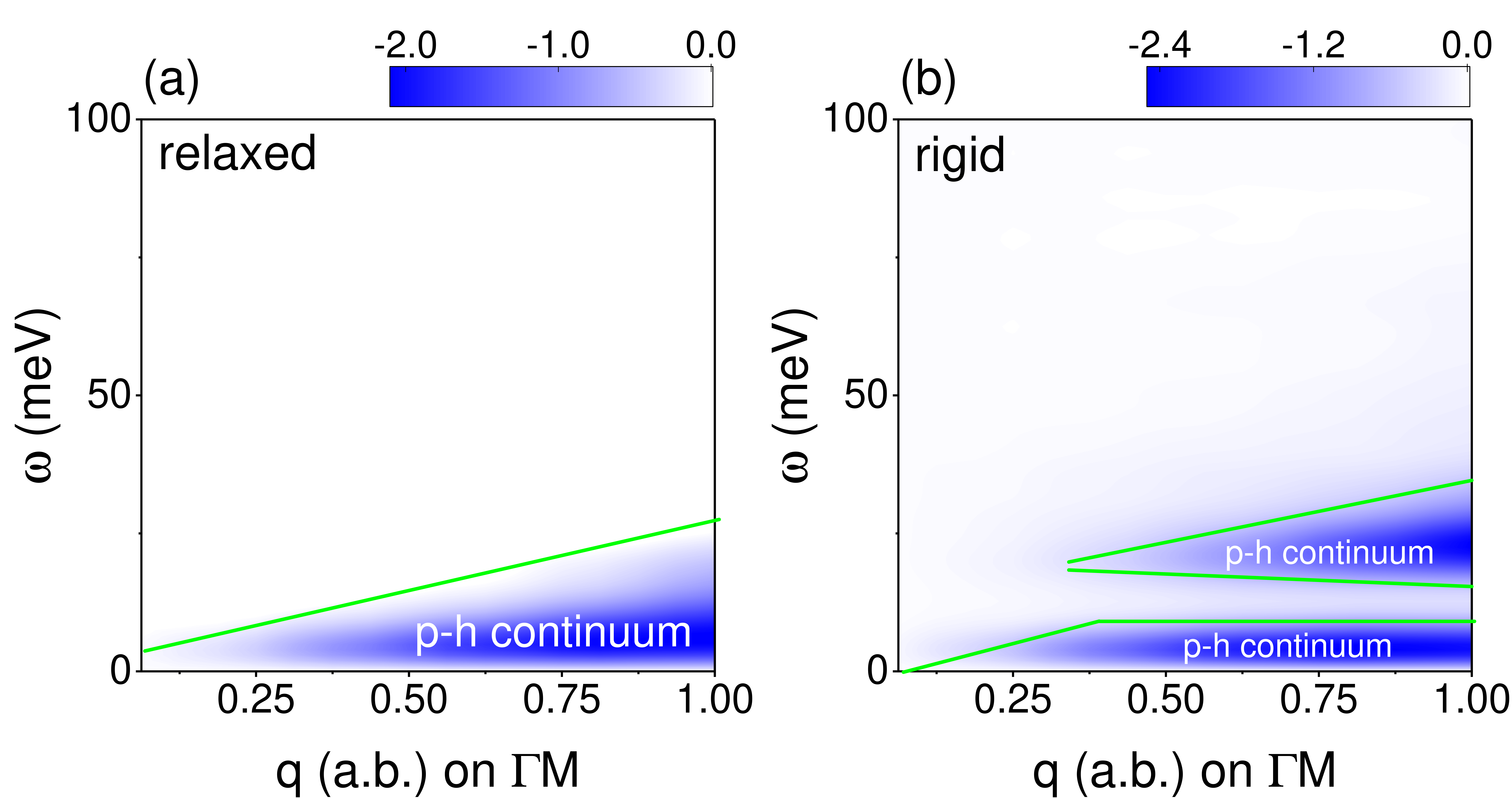}
	\caption{Particle-hole continuum spectra (Im $\Pi(\textbf{q},\omega)$) for (a) relaxed and (b) rigid tb-{\mos} with $3.5^\circ$, in full-band calculation with Eq. (\ref{tbpm}). Green lines are boundaries of non-zero regions in p-h continuum.} 
	\label{fig:phc}
\end{figure}
The main text shows the p-h continuum region and its boundaries in EL function spectra to see if plasmons are subject to Landau damping. Those regions and boundaries are obtained from p-h continuum spectra by calculating Im$\Pi(\textbf{q},\omega)$. Here we show two p-h continuum spectra for relaxed and rigid tb-{\mos} with $3.5^\circ$ in Figs. \ref{fig:phc}(a) and \ref{fig:phc}(b), respectively, obtained from full-band calculations via Eq. (\ref{tbpm}). The continuum spectrum in the rigid case has an energy gap, also shown in Fig. \ref{fig: rigidplas}(a). The continuum region below the gap is the intraband p-h continuum, and the one above the gap is the interband p-h continuum.

\section{Analysis of the polarization function}
\label{app-Pi}
In order to better understand the plasmon behaviors, we display some analytical expressions of the polarization function in this part. \red{The} analytical expressions of the polarization function in a finite-bandwidth two-dimensional electron gas (FBW-2DEG) were studied before in Ref. \onlinecite{plas2020isolate}, in which an electronic energy dispersion is assumed to have the form $E_\mathbf{k} = \hbar^2k^2/2m$ for $k \leq k_c$, and $E_\mathbf{k} = E_c$ otherwise. In the long wavelength limit, $\mathbf{q} \rightarrow 0$, \red{the} real part of the polarization function in FBW-2DEG is
\begin{equation}\label{square}
	\Pi(\textbf{q}, \omega)  \approx \frac{nq^2}{m\omega^2},
\end{equation} 
where $n$ is the charge density and $m$ is the effective mass of charge. By substituting Eq. (\ref{square}) into Eq. (\ref{dielectric}) and letting the dielectric function equal to zero,  we obtain the plasmon dispersion as
\begin{equation}
	\omega_p = \sqrt{\frac{2\pi n e^2q}{m\varepsilon_\mathrm{B}}}
	\label{w2d}.
\end{equation} 
The plasmon modes show a square root of $q$ relation. The Eq. (\ref{w2d}) is used to figure out \red{the} plasmon behavior when $\textbf{q} \rightarrow 0$ in \red{the} 1b-cut calculation in sec.~\ref{relax3.5}.

When $q \geq k_c+k_F$, \red{the} real part of the polarization function is independent of $q$\cite{plas2020isolate}
\begin{equation}\label{pf-flat}
	\text{Re} \Pi(\textbf{q}, \omega) = \frac{m}{\pi\hbar^2}\ln[\frac{E_c^2-\omega^2}{(E_c-E_F)^2-\omega^2}],
\end{equation}
where $E_F$ is the Fermi energy. By inserting Eq. (\ref{pf-flat}) into Eq. (\ref{dielectric}), the plasmon energy dispersion can be derived as
\begin{equation}
	\omega_p =  \sqrt{\frac{E_F(2E_c -E_F)}{exp(q/q_{TF}) -1} + E_c^2},
\end{equation} 
where $q_{TF}$ is the Thomas-Fermi vector. This analytical relation can be used to fit the quasi-flat plasmons in sec.~\ref{relax3.5}, with $|E_c|$ as an effective bandwidth of the flat band.

Next, we will display the expression of $B(\textbf{q})$ and $A(\textbf{q})$ in Eq. (\ref{BqAq}). The polarization function can be divided into two parts in terms of the difference between band transition energy $ |E_{\mathbf{k'} l'} - E_{\mathbf{k}l}|$ and $\omega$\cite{pnasplas2019intrinsically}. 
After considering a $\textbf{k} \rightarrow -\textbf{k}$ time-reversal symmetry replacement in Eq. (\ref{Lindhard}), the polarization function is\cite{pnasplas2019intrinsically}
\begin{equation}\label{lindpnas}
	\begin{aligned}
		\Pi(\textbf{q}, \omega) = &2\frac{g_s}{(2\pi)^2}\int_\mathrm{BZ}d^2\textbf{k}\sum_{l,l'}\frac{n_\mathrm{F}(E_{\mathbf{k}l})F_{\mathbf{k',k}}^{l'l}(E_{\mathbf{k}l}-E_{\mathbf{k'} l'})}
		{(E_{\mathbf{k'} l'} - E_{\mathbf{k}l})^2-(\omega+\mathrm{i}0)^2},
	\end{aligned}
\end{equation}
where $F_{\mathbf{k',k}}^{l'l}$ is the band coherence factor $|\langle \mathbf{k'} l'|\mathrm e^{\mathrm{i}\mathbf{q\cdot r}}|\mathbf{k}l \rangle |^2$ in Eq. (\ref{Lindhard}). \red{The} real part of the polarization 
function Eq. (\ref{lindpnas}) corresponding to those relative high-energy transitions $|E_{\mathbf{k'} l'} - E_{\mathbf{k}l}| > \omega$, gives 
\begin{equation}\label{highener}
	\begin{aligned}
		\Pi_A(\textbf{q}) \approx &\frac{-2g_s}{(2\pi)^2}\int_\mathrm{BZ}d^2\textbf{k}\sum_{l,l'}{^{'}} {F_{\mathbf{k',k}}^{l'l}\frac{n_\mathrm{F}(E_{\mathbf{k}l})}{E_{\mathbf{k'} l' }- E_{\mathbf{k}l}}},
	\end{aligned}
\end{equation}
and \red{the} low-energy transition part $ |E_{\mathbf{k'} l'} - E_{\mathbf{k}l}| <\omega$ is
\begin{equation}\label{lowener}
	\begin{aligned}
		\Pi_B(\textbf{q}, \omega) \approx &\frac{2g_s}{(2\pi)^2\omega^2}\int_\mathrm{BZ}d^2\textbf{k}\sum_{l,l'}{^{''}} F_{\mathbf{k',k}}^{l'l}n_\mathrm{F}(E_{\mathbf{k}l})(E_{\mathbf{k'} l' - E_{\mathbf{k}l}}).
	\end{aligned}
\end{equation}
Note that the summation $\sum_{l,l'}{^{'}}$ and $\sum_{l,l'}{^{''}}$ run over the band indices satisfying $|E_{\mathbf{k'} l'} - E_{\mathbf{k}l}| > \omega $ and $|E_{\mathbf{k'} l'} - E_{\mathbf{k}l}| < \omega$, respectively.
Then we can get an approximate dielectric function
\begin{equation}
	\varepsilon(\textbf{q}, \omega) \approx 1 + A(\textbf{q}) - \frac{B(\textbf{q})}{\omega^2},
	\label{dielectric2}
\end{equation}
by replacing $\Pi(\textbf{q}, \omega) \approx \Pi_A(\textbf{q}, \omega) + \Pi_B(\textbf{q}, \omega) $ in Eq. (\ref{dielectric}), where $A(\textbf{q})$ and $B(\textbf{q})$ are defined as $A(\textbf{q}) = -V(q)\Pi_A(\textbf{q})$ and $B(\textbf{q}) = \omega^2V(q)\Pi_B (\textbf{q},\omega)$. As a consequence, the plasmon energy expression can be written as
\begin{equation}
	\omega_p ^2 \approx \frac{B(\textbf{q})}{1 + A(\textbf{q})}.
\end{equation}
It is obvious that the plasmon energy $\omega_p$ can be further affected by extra band transitions included in $A(\textbf{q})$ or $B(\textbf{q})$, which leads to $\omega_p$ getting lower or higher, as explained in sec.~\ref{relax3.5} and Ref. \onlinecite{pnasplas2019intrinsically}.

\section{Energy loss functions for relaxed and rigid tb-{\mos} with $3.5^\circ$}
\label{app-lossqu}
\begin{figure}[h!] 
	\includegraphics[width=0.5\textwidth]{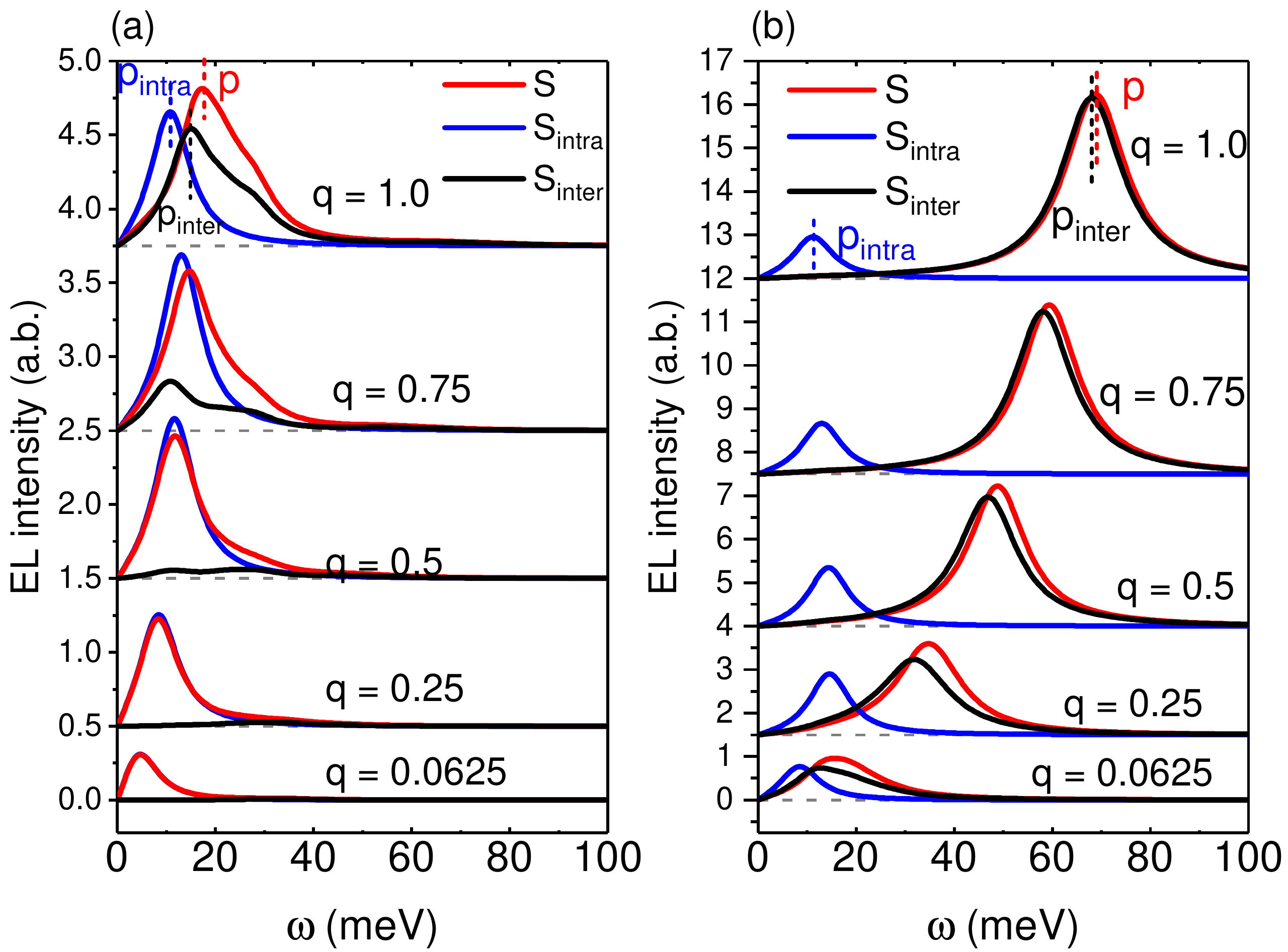}
	\caption{EL functions $S$ (red lines), intraband EL functions $S_{intra}$ (blue lines), and intrerband EL functions $S_{inter}$ (black lines) at five sampled momenta q with unit $\gm$ for $3.5^\circ$ relaxed tb-{\mos} when (a) $\mu$ = -0.4 meV, and (b) $\mu$ = -4.4 meV under 20b-cut calculations.} 
	\label{fig:relaxmu5q}
\end{figure}

\begin{figure}[h!]
	\includegraphics[width=0.5\textwidth]{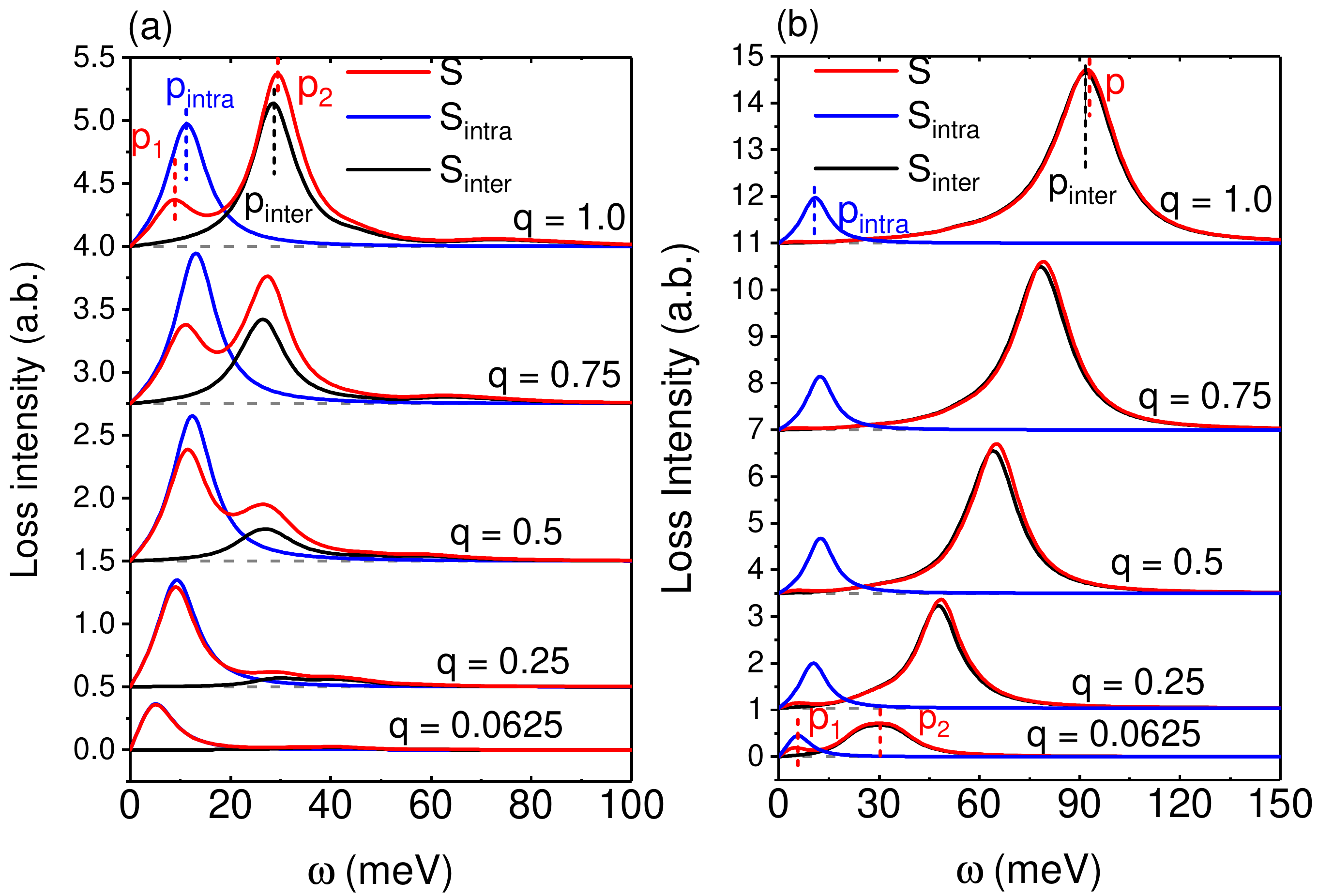}
	\caption{EL functions $S$ (red lines), intraband EL functions $S_{intra}$ (blue lines), and intrerband EL functions $S_{inter}$ (black lines) at five sampled momenta q with unit $\gm$, for $3.5^\circ$ rigid tb-{\mos} when (a) $\mu$ = -0.5 meV, and (b) $\mu$ = -4.0 meV, under 20b-cut calculations. There are two plasmon modes $p_1$ and $p_2$ when q = 0.0625$|\Gamma M|$ in (b).} 
	\label{fig:rigidmu5q}
\end{figure}
We complement EL functions, intraband EL functions, and interband EL functions  in Fig. \ref{fig:relaxmu5q} and Fig. \ref{fig:rigidmu5q} for relaxed and rigid tb-{\mos} with $3.5^\circ$. The EL functions in Figs. \ref{fig:relaxmu5q}(a) with $\mu$ = -0.4 meV and \ref{fig:relaxmu5q}(b) with $\mu$ = -4.4 meV correspond to those plasmon energies in Fig. \ref{fig:mucutcom} (d). For the low hole-doping level in Fig. \ref{fig:relaxmu5q}(a), the intraband plasmon mode $p_{intra}$ gives the dominant contribution to the plasmon $p$ when $q < 1.0\gm$, due to the weak interband transitions for most of q. For $q = 1.0\gm$, \red{the} plasmon mode $p$ is contributed by both intraband and interband plasmons, for enhanced interband transitions at the largest q. However, for the high hole-doping level in Fig. \ref{fig:relaxmu5q}(b), the plasmon mode $p$ is mainly contributed by $p_{inter}$, with an obvious overlap between $S_{inter}$ and $S$ even when $q=0.0625\gm$. \red{The} slightly larger plasmon energy of $p$ than $p_{inter}$ arises from the enhanced $B(\textbf{q})$ term by extra low energy intraband transitions.
Comparing \red{the} interband EL function $S_{inter}$ at $\mu$ = -0.4 meV  to $\mu$ = -4.4 meV, we can verify that interband transitions are enhanced with \red{a} deep doping, as shown in Fig. \ref{fig:muloss}(a) with q = 0.5$\gm$.

In the rigid tb-{\mos}, when the flat band is slightly filled ($\mu$ = -0.5 meV) in Fig. \ref{fig:rigidmu5q}(a), the quasi-flat plasmons $p_1$ contributed by \red{the} intraband plasmons $p_{intra}$ are stronger than $p_2$ from q = 0.0625$\gm$ to q = 0.5$\gm$ and then gradually weakened by \red{the} higher-energy interband transitions. The higher-energy plasmons $p_2$ appear from q = 0.5$\gm$ and are gradually strengthened with q. When the flat band is almost fully filled ($\mu$ = -4.0 meV) in Fig. \ref{fig:rigidmu5q}(b), we observe \red{the} two plasmon modes $p_1$ and $p_2$ only for the sampled smallest q = 0.0625$\gm$. For other momenta, there is only one plasmon mode $p$ dominated by higher-energy interband plasmon $p_{inter}$ as strong interband transitions thoroughly screen the $p_{intra}$. Comparing intraband EL functions $S_{intra}$ at $\mu$ = -4.0 meV with $\mu$ = -0.5 meV, $S_{intra}$ have similar intensity in the two cases, and $p_{intra}$ always exist at $\mu$ = -4.0 meV. Therefore, we can verify that the vanishment of \red{the} quasi-flat plasmons in Fig. \ref{fig:rigidmuplas}(a) is caused by the enhanced higher-energy interband transitions.

\section{Interband coherence and Fermi-Dirac factor in polarization function}
\label{app-corre}
\begin{figure}[h!]
	\includegraphics[width=0.5\textwidth]{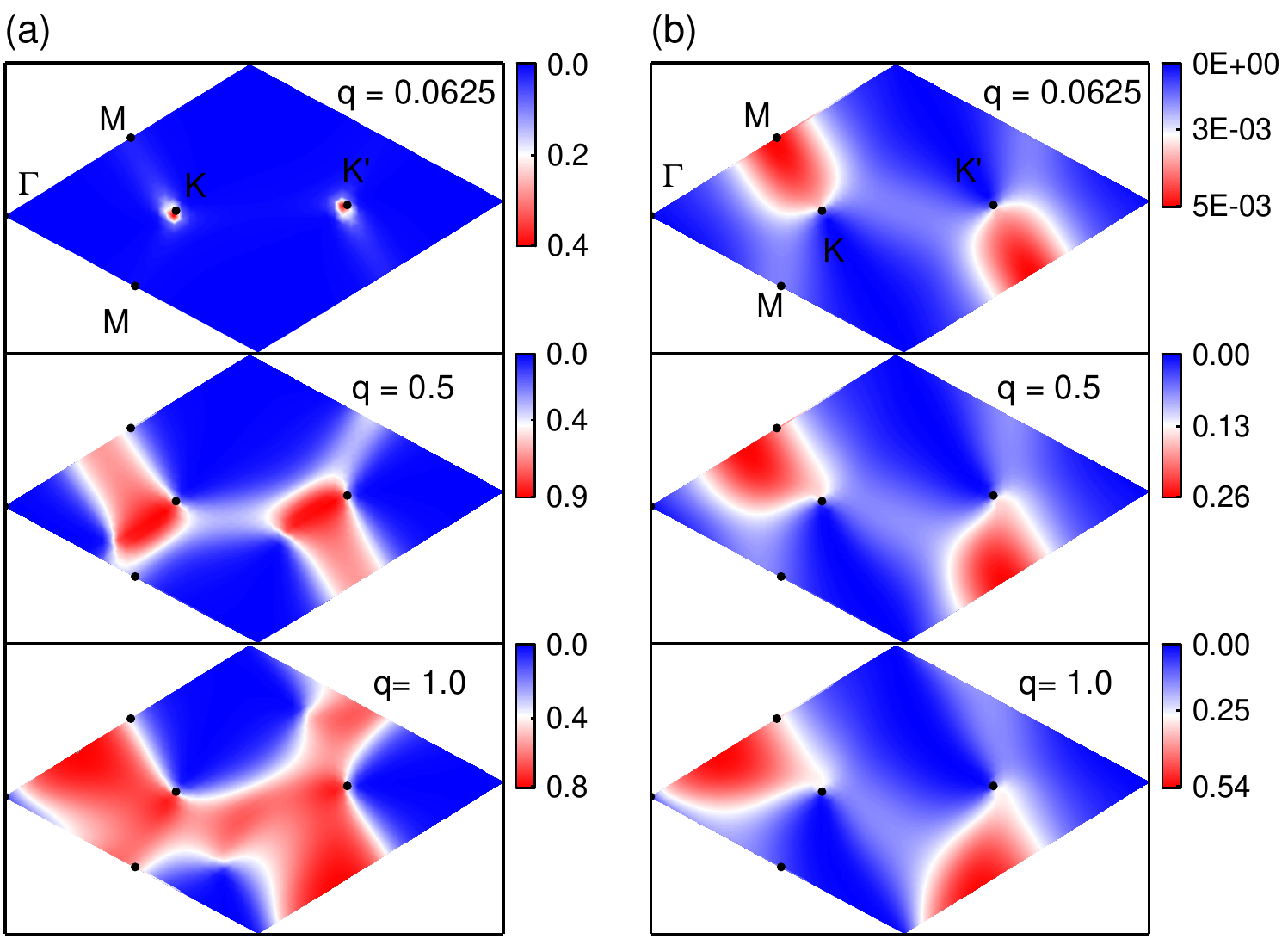}
	\caption{Intensity plots of interband coherence factor $F_{\mathbf{k',k}}^{19,20}$ between the doped flat band and its nearest-neighbor valence band, for (a) relaxed and (b) rigid tb-{\mos} with $3.5^\circ$. Upper, middle and bottom panels in (a) and (b) represent different length of momenta with  q = 0.0625$|\Gamma M|$, q = 0.5$|\Gamma M|$, and q = 1.0$|\Gamma M|$, respectively.} 
	\label{fig:corre}
\end{figure}

In this part, we will analyze \red{the} interband contrbution to plasmons at different momenta q and 
chemical potentials $\mu$ in relaxed and rigid tb-{\mos}. The interband contribution can be affected by the band coherence factor $F_{\mathbf{k',k}}^{l'l} = |\langle \mathbf{k'} l'|\mathrm e^{\mathrm{i}\mathbf{q\cdot r}}|\mathbf{k}l \rangle |^2$ with different momenta and \red{the} Fermi-Dirac factor $f_{\mathbf{k',k}}^{l'l} = n_\mathrm{F}(E_{\mathbf{k'} l'}) - n_\mathrm{F}(E_{\mathbf{k}l})$ with various chemical potentials in Eq. \ref{Lindhard}. Here, we focus on the interband coherence factor and Fermi-Dirac factor between the doped flat band and its nearest-neighbor VB, \red{the} band indices of which are denoted by $l = 20$ and $l'= 19$, respectively, in \red{the} 20b-cut calculation. The band coherence factor $F_{\mathbf{k',k}}^{19,20}$ can imply the interband correlation weighted by  $f_{\mathbf{k',k}}^{19,20}$. The spectra of $F_{\mathbf{k',k}}^{19,20}$ at \red{the} three sampled q are displayed in Figs. \ref{fig:corre} (a) and \ref{fig:corre} (b) for relaxed and rigid $3.5^\circ$  tb-{\mos}, respectively. The region of non-zero $F_{\mathbf{k',k}}^{19,20}$ in relaxed $3.5^\circ$  tb-{\mos} changes a lot with q, compared to much smaller variation of the non-zero area in the rigid case. A broader area of non zeros will give more interband contribution to Eq. (\ref{Lindhard}), and thus enhance interband EL functions with \red{a} larger q, as shown in Fig. \ref{fig:relaxloss}(a) and Fig. \ref{fig:relaxmu5q}. For \red{the} rigid case, the intensity of $F_{\mathbf{k',k}}^{19,20}$ increases a lot with \red{a} larger q. For example, the maximum of $F_{\mathbf{k',k}}^{19,20}$ for q at 0.5$\gm$ is hundreds of times larger than the one at q = 0.0625$\gm$, although the change of non-zero zone is not so significant as the relaxed case from q = 0.0625$\gm$ to 0.5$\gm$. As a result, the interband EL functions shown in Fig. \ref{fig: relaxplas} (b) and Fig. \ref{fig:rigidmu5q} can be mainly enhanced by more significant intensity of  $F_{\mathbf{k',k}}^{l'l}$. We remark that the enhanced interband coherence factor and its more non-zero terms in momenta space could increase \red{the} interband contribution to polarization function, which leads to \red{the} enhanced interband plasmons with \red{a} higher energy.

\begin{figure}[t!]
	\includegraphics[width=0.5\textwidth]{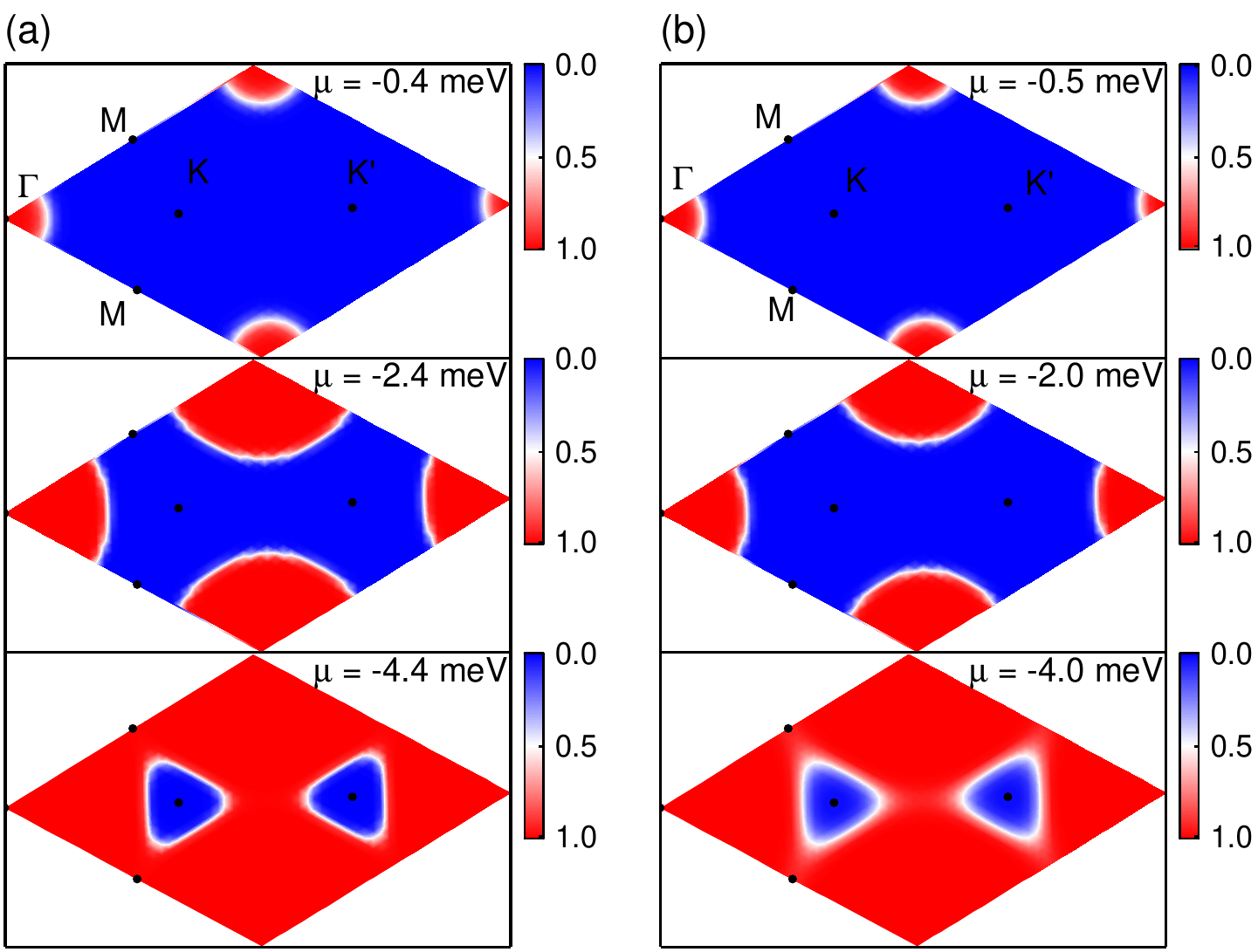}
	\caption{Intensity plots of Fermi-Dirac factor $f_{\mathbf{k',k}}^{19,20}$ between the doped flat band and its nearest-neighbor valence band, for (a) relaxed and (b) rigid tb-{\mos} with $\theta=3.5^\circ$ and q = 0.5$\gm$. Upper, middle and bottom panel represents the intensity of plots at $\mu$ = -0.4 meV, $\mu$ = -2.4 meV and $\mu$ = -4.4 meV in (a), and at  $\mu$ = -0.5 meV, $\mu$ = -2.0 meV and $\mu$ = -4.0 meV in (b).}
	\label{fig:fermi}
\end{figure}

We have displayed that tuning chemical potentials $\mu$ also changes the interband EL functions a lot, seen in Fig. \ref{fig:muloss}, through affecting Fermi-Dirac factors $f_{\mathbf{k',k}}^{l'l}$ in Eq. (\ref{Lindhard}). The intensity plots of $f_{\mathbf{k',k}}^{19,20}$ at three chemical potentials are shown in Figs. \ref{fig:fermi} (a) and \ref{fig:fermi} (b) for relaxed and rigid $3.5^\circ$ tb-{\mos}, respectively, with q = 0.5$\gm$. In both cases, The non-zero area of $f_{\mathbf{k',k}}^{19,20}$ broadens with larger doping levels, showing more interband transitions over the BZ can contribute to the polarization function and thus enhance the interband EL functions and interband plasmons in Fig. \ref{fig:muloss}. The enhanced interband transitions can be easily understood as more holes occupied in the flat band at a large $|\mu|$ leading to more possible electron-hole interband transitions. The non-zero terms are determined by the hole-occupied part of \red{the} flat band over the BZ. The boundaries of non-zero $f_{\mathbf{k',k}}^{19,20}$ are denoted by the white and bright spectrum weight (intensity around 0.5) in Figs. \ref{fig:fermi}(a) and \ref{fig:fermi}(b).

\bibliographystyle{apsrev4-1}
\bibliography{reference}

\end{document}